\documentclass[12pt]{article}

\usepackage{a4}
\usepackage{latexsym}
\usepackage{cite}
\usepackage{url}
\usepackage{hyperref}

\usepackage{color}
\usepackage{xcolor}

\usepackage{array}
\newcolumntype{T}{>{\ttfamily} c}
\newcolumntype{M}{>{$\displaystyle} c <{$}}

\usepackage{epsfig}
\usepackage{graphicx}
\usepackage{amsmath}
\usepackage{amssymb}
\usepackage{multirow}
\allowdisplaybreaks

\usepackage{pslatex}
\usepackage[latin1]{inputenc}
\usepackage[T1]{fontenc}

\usepackage{color,colordvi}
\def\colour4colour#1{\Blue{#1}}

\usepackage{graphicx}
\usepackage{float}
\usepackage{todonotes}
\usepackage{breqn}
\usepackage{slashed}

\newcommand{\eps}{\varepsilon}
\newcommand{\Op}{\mathcal{O}}

\newcommand{\eom}{\text{EOM }}

\newcommand{\MSb}{$\overline{\mbox{MS}}$ }

\def\gs{{g_{s}}}
\def\as(#1){{\alpha_{s}^{\,#1}}}
\def\gss{{g_{s}^{2}}}

\numberwithin{equation}{section}

\begin{document}
\setlength{\parskip}{0.2cm}
\setlength{\baselineskip}{0.54cm}


\begin{titlepage}
\noindent
ZU-TH 43/24 \hfill August 2024 \\
DESY-24-108 \\
\vspace{0.6cm}
\begin{center}
{\LARGE \bf Constraints for twist-two alien operators in QCD
}\\
\vspace{2.0cm}
\large
G. Falcioni$^{\, a,b}$,
F. Herzog$^{\, c}$,
S. Moch$^{\, d}$ and
S. Van Thurenhout$^{\, e}$
\\

\vspace{1.2cm}
\normalsize
{\it $^a$ Dipartimento di Fisica, Universit\`a di Torino, Via Pietro Giuria 1, 10125 Torino, Italy}\\
\vspace{4mm}
{\it $^b$ Physik-Institut, Universit\"at Z\"urich, Winterthurerstrasse 190, 8057 Z\"urich, Switzerland}\\
\vspace{4mm}
{\it $^c$ Higgs Centre for Theoretical Physics, School of Physics and Astronomy, \\ \vspace{0.5mm}
The University of Edinburgh, Edinburgh EH9 3FD, Scotland, UK}\\
\vspace{4mm}
{\it $^d$II.~Institute for Theoretical Physics, Hamburg University\\
\vspace{0.5mm}
Luruper Chaussee 149, D-22761 Hamburg, Germany}\\
\vspace{4mm}
{\it $^e$ HUN-REN Wigner Research Centre for Physics, Konkoly-Thege Mikl\'os u. 29-33, 1121
Budapest, Hungary}\\
\vspace{2.4cm}
{\large \bf Abstract}
\vspace{-0.2cm}
\end{center}
\vspace*{0.5cm}

Parton evolution equations in QCD are controlled by the anomalous dimensions of gauge-invariant twist-two spin-$N$ quark and gluon operators.
Under renormalization, these mix with gauge-variant operators of the same quantum numbers, referred to as alien operators.
Our work addresses the systematic study of these alien operators at arbitrary spin $N$, using generalized BRST symmetry relations to derive their couplings and Feynman rules at all values of $N$.
We observe how the all-$N$ structure of the generalized (anti-)BRST constraints relates the couplings of alien operators with $n+1$ gluons to those with $n$ gluons. 
Realizing a bootstrap, we present all one-loop results necessary for performing the operator renormalization up to four loops in QCD.

\end{titlepage}

\newpage

\section{Introduction}
\label{sec:intro}

The study of twist-two operators of spin-$N$ for quarks and gluons in quantum chromodynamics (QCD) and their renormalization dates to the origins of QCD as the gauge theory of the strong interaction~\cite{Dixon:1974ss,Kluberg-Stern:1974nmx,Kluberg-Stern:1975ebk,Joglekar:1975nu,Joglekar:1976eb,Joglekar:1976pe}. 
The renormalization of off-shell operator matrix elements (OMEs) in QCD, i.e. Green's functions with off-shell external momenta and insertions of these quark and gluon operators, gives access to their anomalous dimensions.
These coincide with the Mellin transforms of the standard QCD splitting functions, that govern the scale evolution of the parton distribution functions.
It is well-known that the twist-two operators of spin-$N$ mix under renormalization
with a set of gauge-variant operators of the same quantum numbers, which
involve equation-of-motion (EOM) and ghost operators.
The latter, often referred to in summary as \textit{alien} operators, can be constructed systematically, by employing a generalized gauge symmetry of the QCD Lagrangian in covariant gauge with the addition of the physical quark and gluon operators \cite{Dixon:1974ss,Joglekar:1975nu,Hamberg:1991qt,Falcioni:2022fdm}.
The generalized gauge symmetry can be promoted to a generalized BRST (gBRST) symmetry \cite{Joglekar:1975nu,Falcioni:2022fdm}. 
This provides an algebraic approach for the derivation of a complete set of operators to be considered in the renormalization of the off-shell OMEs at a given loop order in perturbative QCD in an expansion in the strong coupling $g_s$, $\alpha_s=g_s^2/(4\pi)$. The complete set of operators required up to four loops has been listed in~\cite{Falcioni:2022fdm,Falcioni:2024xyt}.

Each alien operator features a coupling constant that can be interpreted as the renormalization constant that generates mixing of the gauge-invariant operators into each alien. In order to renormalize the physical OMEs, these coupling constants must be computed order-by-order in perturbation theory. The required couplings to renormalize the two-loop OMEs were computed in~\cite{Dixon:1974ss,Hamberg:1991qt} in closed form for all values of $N$. A method to determine the alien counterterms, i.e. the Feynman rules obtained by summing all the alien operators with their associated couplings, was presented in~\cite{Gehrmann:2023ksf} together with results up to the three-loop level for a covariant gauge and all values of $N$. From this, the $n_f^2$ contributions to the pure-singlet splitting functions were obtained at four loops~\cite{Gehrmann:2023cqm}. Beyond three loops, ref.~\cite{Falcioni:2022fdm} determined a set of all-order constraints on the couplings, induced by gBRST and generalized anti-BRST symmetries \cite{Curci:1976bt,Ojima:1980da,Baulieu:1981sb}. In \cite{Falcioni:2022fdm,Falcioni:2024xyt}, these constraints were solved at arbitrary loop order for fixed $N\leq20$, leaving the systematic study of the alien operators at arbitrary spin $N$ as an open problem. In this paper, we follow a different strategy. Namely, we will solve the constraints on the alien couplings to leading order in $\alpha_s$ but for \textit{all} values of $N$. The main results of our study are:

\begin{itemize}
\item The all-$N$ structure of the couplings is fixed in terms of a small set of constants. The latter can be determined by explicitly computing the couplings for some fixed values of $N$.
\item The structure of the couplings of alien operators with $n+1$ gluons is related to the ones with $n$ gluons, allowing for a bootstrap in the determination of complicated higher-order couplings in terms of simpler lower-order ones. 
\end{itemize}
The outline of the article is as follows. 
In Sec.~\ref{sec:set-stage} we set the stage, review the generalized gauge symmetry and provide a brief summary of the set of relevant alien operators. In Sec.~\ref{sec:identities} we study the identities that exist among the couplings of the alien operators and show how they can be used to restrict the all-$N$ structure of the couplings. The results of this analysis are then used in Sec.~\ref{sec:feynman} to derive the Feynman rules of the alien operators, suitable for the renormalization of OMEs at all $N$ up to four loops in QCD. Finally, in Sec.~\ref{sec:concl}, we summarize our findings and provide an outlook on further developments.

\section{Setting the stage}
\label{sec:set-stage}
In this section, we review the construction of the alien operators and summarize our conventions. The complete gauge-fixed QCD action is written as
\begin{equation}
    S=
    \int \text{d}^{D}x\, 
    \left(\mathcal{L}_0+\mathcal{L}_{\mathrm{GF+G}}\right)\,.
\end{equation}
Here $\mathcal{L}_{0}$ represents the classical part of the QCD Lagrangian
\begin{equation}
\label{eq:L0}
    \mathcal{L}_{0} = -\frac{1}{4}\,F^{\mu\nu}_a\,F_{\mu\nu}^{a} + \sum_{f=1}^{n_f} \overline{\psi}^f(i\slashed{D}-m_f)\psi^f\,,
\end{equation}
with the field strength defined as
\begin{equation}
  F_{\mu\nu}^a = \partial_\mu A_\nu^a - \partial_\nu A_\mu^a + g_s f^{abc} A_\mu^bA_\nu^c\,.
\end{equation} 
$f^{abc}$ are the standard QCD structure constants. The covariant derivative in Eq.~(\ref{eq:L0}) is $\slashed{D}= \gamma^\mu(\partial_\mu-ig_s T^aA^a_\mu)$ with $T^a$ the generator of the gauge group in the fundamental representation. The gauge-fixing and ghost terms are
\begin{equation}
  \label{def:GF}
  \mathcal{L}_{\text{GF+G}}=-\frac{1}{2\xi}(\partial^\mu A^{a}_\mu)^2-\overline{c}^a\,\partial^\mu D_\mu^{ab}\,c^b\,
\end{equation}
with $\xi$ the covariant gauge parameter and $\overline{c}^a$ and $c^a$ the anti-ghost and ghost fields, respectively. The covariant derivative in the adjoint representation is $D_\mu^{ac}=\partial_{\mu}\delta^{ac}+g_s f^{abc}A_{\mu}^{b}$. The QCD Lagrangian can be extended to also include spin-$N$ gauge-invariant operators of twist two, which we define as
\begin{equation}
\begin{aligned}
\label{def:GIop}
\Op_{\rm g}^{(N)}(x) &=& 
\frac{1}{2}\text{Tr}\big[F_{\nu}(x)\,D^{N-2} F^{\nu}(x)\big]\,,\\
\Op_{\rm q}^{(N)}(x) &=& \frac{1}{2}\text{Tr}\big[\overline{\psi}(x)\slashed{\Delta}\:D^{N-1}\psi(x)\big] \,.
\end{aligned}
\end{equation}
Here $\Delta_\mu$ is a lightlike vector and we introduced the notation
\begin{align}
  F^{\mu;a}=\Delta_\nu\,F^{\mu\nu;a},\qquad  A^a=\Delta_\mu A^{\mu;a},\qquad D=\Delta_\mu\,D^\mu,\qquad \partial=\Delta_\mu\partial^\mu\,.
\end{align}
Under renormalization the operators in Eq.~(\ref{def:GIop}) mix with operators proportional to the (classical) EOM and with BRST-exact operators \cite{Dixon:1974ss,Joglekar:1975nu}. Following \cite{Falcioni:2022fdm,Falcioni:2024xyt}, we begin by presenting the EOM aliens in the form
\begin{equation}
\label{eq:EOMdefG}
 \Op_{\text{EOM}}^{(N)}=\left(D\cdot F^a  +g_s\overline{\psi}T^a\slashed{\Delta}\psi\right) \mathcal{G}^a(A^a,\partial A^a,\partial^2  A^a,... )\,
\end{equation}
with $D\cdot F^a = D_\nu F^{\nu;a}$ and $\mathcal{G}^a$ a generic local function of the gauge field and its derivatives. It is convenient to expand $\mathcal{G}^a$ in a series of contributions with an increasing number of gauge fields. This leads to
\begin{equation}
\label{eq:eom-FULL}
    \Op_{\eom}^{(N)} = \Op_{\eom}^{(N),I}+\Op_{\eom}^{(N),II}+\Op_{\eom}^{(N),III}+\Op_{\eom}^{(N),IV}+\:\dots
\end{equation}
with
\begin{align}
\label{eq:eomI}
 \Op_{\text{EOM}}^{(N),I} &= 
    \eta(N)\, \left(D\cdot F^a + g_s \overline{\psi}\,\slashed\Delta\,T^a 
    \psi\right)\, \left(\partial^{\,N-2}A^a \right), \\
    \label{eq:eomII}
 \Op_{\text{EOM}}^{(N),II} &= 
 \gs \left(D\cdot F^a + \gs \overline{\psi}\,\slashed \Delta\,
   T^a \psi\right) \sum_{\substack{i+j\\=N-3}}C_{ij}^{abc}(\partial^{i}A^{b})(\partial^{j}A^{c}),\\
   \label{eq:eomIII}
 \Op_{\text{EOM}}^{(N),III} &= 
  \gss \,
  \left(D\cdot F^a + \gs \overline{\psi}\slashed\Delta T^a\psi\right) \sum_{\substack{i+j+k\\=N-4}}C_{ijk}^{abcd}(\partial^{i}A^{b})(\partial^{j}A^{c})(\partial^{k}A^{d}),\\
  \label{eq:eomIV}
  \Op_{\text{EOM}}^{(N),IV} &= 
  g_s^3 \,
  \left(D\cdot F^a + \gs \overline{\psi}\slashed\Delta T^a\psi\right)\sum_{\substack{i+j+k+l\\=N-5}}C_{ijkl}^{abcde}(\partial^{i}A^{b})(\partial^{j}A^{c})(\partial^{k}A^{d})(\partial^{l}A^{e}).
\end{align}
The coefficients $C^{a_1\dots a_n}_{i_1\dots i_{n-1}}$ appearing in Eqs.~(\ref{eq:eomII})-(\ref{eq:eomIV}) can be written in terms of a set of independent colour tensors, each of them multiplying an associated coupling constant, as follows
\begin{align}
    &C_{ij}^{abc} = f^{abc}\kappa_{ij}, \label{def:Cij}\\
    &C_{ijk}^{abcd}=(f\: f)^{abcd}\kappa_{ijk}^{(1)}+d_{4}^{abcd}\kappa_{ijk}^{(2)}+d_{\widehat{4ff}}^{abcd}\kappa_{ijk}^{(3)}, \label{eq:Cijk}\\
    &C_{ijkl}^{abcde} = (f\:f\:f)^{abcde}\kappa_{ijkl}^{(1)}+d_{4f}^{abcde}\kappa_{ijkl}^{(2)}\label{def:Cijkl}    
\end{align}
with
\begin{align}
\label{def:ff}
    &(f\: f)^{abcd} = f^{abe}f^{cde}, \\
    &(f\:f\:f)^{abcde} = f^{abm}f^{mcn}f^{nde}, \\
    &d_{4}^{abcd} = \frac{1}{4!}[\text{Tr}(T_{A}^{a}T_{A}^{b}T_{A}^{c}T_{A}^{d})+\text{symmetric permutations}], \\
    &d_{4ff}^{abcd} = d_{4}^{abmn}f^{mce}f^{edn}, \\
    &d_{\widehat{4ff}}^{abcd} = d_{4ff}^{abcd}-\frac{1}{3}C_{A} d_{4}^{abcd}, \\
\label{def:d4f}    
    &d_{4f}^{abcde} = d_{4}^{abcm}f^{mde}.
\end{align}
Here $(T_{A})^{b}_{ac} = i f^{abc}$ are the generators of the adjoint representation of the colour group. We now extend the classical Lagrangian $\mathcal{L}_0$ in Eq.~(\ref{eq:L0}) to include the gauge-invariant operators of twist two as well as the EOM aliens
\begin{equation}
    \mathcal{L}_{\text{GGI}} = \mathcal{L}_0+w_{\rm i}\,\Op_{\rm i}^{(N)}+\Op_{\text{EOM}}^{(N)},
\end{equation}
where $w_{\rm i}$ is a coupling for the operator $\Op_{\rm i}$ with $\rm i=g,q$, playing the same role as the coefficients $\eta(N)$, $\kappa_{ij}$, $\dots$ defined in Eqs.~(\ref{def:Cij})-(\ref{def:Cijkl}). The Lagrangian $\mathcal{L}_{\text{GGI}}$ is invariant under the generalized gauge transformation \cite{Falcioni:2022fdm} $A^a_\mu\to A^a_\mu + \delta_\omega A^a_\mu + \delta_\omega^\Delta A^a_\mu$, where 
\begin{equation}
\begin{aligned}
    \delta_\omega A^a_\mu &= D^{ab}_\mu\omega^b(x),\\
    \label{eq:gengauge}
    \delta_\omega^\Delta A^a_\mu &= -\Delta_\mu\Bigg[\eta(N)\,\partial^{N-1}\omega^a + g_s\sum_{\substack{i+j\\=N-3}} \widetilde{C}^{aa_1a_2}_{ij}\,\left(\partial^{i}A^{a_1}\right)\,\left(\partial^{j+1}\omega^{a_2}\right)\\
    &+g_s^2\sum_{\substack{i+j+k\\=N-4}}\widetilde{C}^{aa_1a_2a_3}_{ijk}\,\left(\partial^{i}A^{a_1}\right)\,\left(\partial^{j}A^{a_2}\right)\,\left(\partial^{k+1}\omega^{a_3}\right)\\
    &+g_s^3\,\sum_{\substack{i+j+k+l\\=N-5}}\widetilde{C}^{aa_1a_2a_3a_4}_{ijkl}\,\left(\partial^{i}A^{a_1}\right)\,\left(\partial^{j}A^{a_2}\right)\,\left(\partial^{k}A^{a_3}\right)\,\left(\partial^{l+1}\omega^{a_4}\right) + {\cal{O}}(g_s^4)\Bigg]
\end{aligned}
\end{equation}
and 
\begin{align}
    &\widetilde{C}_{ij}^{abc} = f^{abc}\eta_{ij},  \\
    &\widetilde{C}_{ijk}^{abcd}=(f \: f)^{abcd}\eta_{ijk}^{(1)}+d_{4}^{abcd}\eta_{ijk}^{(2)}+d_{\widehat{4ff}}^{abcd}\eta_{ijk}^{(3)}, \\
    &\widetilde{C}_{ijkl}^{abcde} = (f\:f\:f)^{abcde}\eta_{ijkl}^{(1)}+d_{4f}^{abcde}\eta_{ijkl}^{(2a)}+d_{4f}^{aebcd}\eta_{ijkl}^{(2b)}.
\end{align}
The generalized gauge symmetry implies that the couplings $\eta^{(k)}_{n_1\dots n_j}$ are related to $\kappa^{(k)}_{n_1\dots n_j}$ in Eqs.~(\ref{eq:eomI})-(\ref{eq:eomIV})
\begin{align}
\label{def:etaij}
    \eta_{ij}&=2\kappa_{ij}+\eta(N)\binom{i+j+1}{i},\\
    \label{def:eta1ijk}
    \eta^{(1)}_{ijk}&=2\kappa_{i(j+k+1)}\binom{j+k+1}{j}+2[\kappa_{ijk}^{(1)}+\kappa_{kji}^{(1)}],\\
    \label{def:eta2ijk}
    \eta^{(2)}_{ijk}&=3\kappa^{(2)}_{ijk},\\
    \label{def:eta3ijk}
    \eta^{(3)}_{ijk}&=2\big[\kappa^{(3)}_{ijk}-\kappa^{(3)}_{kji}\big],\\
    \label{def:eta1ijkl}
    \eta^{(1)}_{ijkl}&=2[\kappa_{ij(l+k+1)}^{(1)}+\kappa_{(l+k+1)ji}^{(1)}]\binom{l+k+1}{k}+2[\kappa_{ijkl}^{(1)}+\kappa_{ilkj}^{(1)}+\kappa_{likj}^{(1)}+\kappa_{lkij}^{(1)}],\\
    \label{def:eta2aijkl}
    \eta^{(2a)}_{ijkl}&=3\kappa_{ij(k+l+1)}^{(2)}\binom{k+l+1}{k}+2\kappa_{ijkl}^{(2)},\\
    \label{def:eta2bijkl}
    \eta^{(2b)}_{ijkl}&=2\kappa^{(2)}_{lijk}.
\end{align}
The new gauge transformations in Eq.~(\ref{eq:gengauge}) are promoted to a nilpotent generalized BRST (gBRST) operator, by replacing the transformation parameter $\omega^a$ with the ghost field $c^a$ \cite{Falcioni:2022fdm}. In turn the ghost alien operator is generated by the action of such gBRST operator on a suitable ancestor operator \cite{Falcioni:2022fdm}, giving
\begin{equation}
\label{eq:ghostFULL}
    \Op_{c}^{(N)} = \Op_{c}^{(N),I}+\Op_{c}^{(N),II}+\Op_{c}^{(N),III}+\Op_{c}^{(N),IV}+\:\dots
\end{equation}
with
\begin{align}
    &\Op_{c}^{(N),I} = -\eta(N) (\partial\overline{c}^{a})(\partial^{N-1}c^{a}), \label{eq:ghost-LO} \\
    &\Op_{c}^{(N),II} = -g_s \sum_{\substack{i+j\\=N-3}}\widetilde{C}_{ij}^{abc}(\partial\overline{c}^{a})(\partial^{i}A^{b})(\partial^{j+1}c^{c}), \label{eq:ghost-NLO} \\
    &\Op_{c}^{(N),III} = -g_s^2\sum_{\substack{i+j+k\\=N-4}}\widetilde{C}_{ijk}^{astu}(\partial\overline{c}^a)(\partial^{i}A^{s})(\partial^{j}A^{t})(\partial^{k+1}c^{u}), \label{eq:ghost-NNLO} \\
    &\Op_{c}^{(N),IV} = -g_s^3\sum_{\substack{i+j+k+l\\=N-5}}\widetilde{C}_{ijkl}^{abcde}(\partial\overline{c}^{a})(\partial^{i}A^{b})(\partial^{j}A^{c})(\partial^{k}A^{d})(\partial^{l+1}c^{e}) \label{eq:ghost-NNNLO}.
\end{align}

\paragraph{Renormalization} The complete Lagrangian, including the twist-two physical and alien operators, can be written as
\begin{align}
\label{eq:LagrNONREN}
   \widetilde{\mathcal{L}} &= \mathcal{L}_0 + \mathcal{L}_{\text{GF+G}} + w_{\rm i}\,\Op_{\rm i} + \Op_{\text{EOM}}^{(N)} + \Op_{c}^{(N)} =\mathcal{L}_0(A^a_\mu,g_s) + \mathcal{L}_{\text{GF+G}}(A^a_\mu,c^a,\bar{c}^a,g_s,\xi) + \sum_{\rm k}\,\mathcal{C}_{\rm k}\,\Op_{\rm k},
\end{align}
where $\mathcal{C}_{\rm k}$ labels all the distinct couplings of the operators, e.g. $\mathcal{C}_{\rm k}=(w_{\rm i},\eta(N), \kappa_{0\,1}, \kappa_{1\,2} \dots)$. The ultraviolet (UV) singularities associated with the QCD Lagrangian are absorbed by introducing the bare fields and parameters
\begin{align}
    &&A^{a;\text{bare}}_\mu(x) = \sqrt{Z_3}\,A^a_\mu(x), &&c^{a;\text{bare}}(x)=\sqrt{Z_c}\,c^a(x), && \bar{c}^{a;\text{bare}}(x)=\sqrt{Z_c}\,\bar{c}^a(x),\\
    &&g_s^{\text{bare}}=\mu^\epsilon\,Z_g g_s, &&\xi^{\text{bare}}=\sqrt{Z_3}\,\xi.
\end{align}
We renormalize the singularities originating from the insertion of the composite operators using
\begin{align}
    \Op_{\rm i}^{\text{ren}}(x) &= Z_{\rm ij}\,\Op_{\rm j}^{\text{bare}}(x),
\end{align}
where $\Op_{\rm j}^{\text{bare}}$ indicates the operators in Eqs.~(\ref{def:GIop}), (\ref{eq:eom-FULL}) and (\ref{eq:ghostFULL}) written in terms of the bare fields. Note that throughout this work we use $D=4-2\eps$ dimensional regularization, combined with the \MSb renormalization scheme. $Z_{\rm ij}$ is the renormalization matrix of the operators, which makes the OMEs featuring an insertion of $\Op_{\rm i}^{\text{ren}}$ finite. The renormalized Lagrangian becomes
\begin{align}
\label{eq:RenLagr}
    &\widetilde{\mathcal{L}} = \mathcal{L}_0(A^{a;\text{bare}}_\mu,g_s^{\text{bare}}) + \mathcal{L}_{\text{GF+G}}(A^{a;\text{bare}}_\mu,c^{a;\text{bare}},\bar{c}^{a;\text{bare}},g_s^{\text{bare}},\xi^{\text{bare}}) + \sum_{\rm k}\mathcal{C}_{\rm k}^{\text{bare}}\,\Op_{\rm k}^{\text{bare}},\\
 &\mathcal{C}_{\rm i}^{\text{bare}} = \sum_{\rm k} \mathcal{C}_{\rm k}\,Z_{\rm k\,i},   
\end{align}
where $\mathcal{C}_{\rm k}$ is the (finite) renormalized coupling of the operator $\Op_{\rm k}$. The UV-finite OMEs featuring a single insertion of $\Op_{\rm g}^{\text{ren}}$ are computed by setting the renormalized couplings $\mathcal{C}_{\rm i}=\delta_{\rm i\,g} $ in Eq.~(\ref{eq:RenLagr}), which gives
\begin{equation}
\mathcal{C}_{\rm i}^{\text{bare}}=Z_{\rm g\,i}.
\end{equation}
Similarly, the renormalized OMEs with an insertion of $\Op_{\rm q}$ are obtained with $\mathcal{C}_{\rm i}^{\text{bare}}=Z_{\rm q\,i}$. Therefore, the couplings of the bare operators $\eta^{\text{bare}}(N)$, $\dots$ are interpreted as the renormalization constants that mix the physical operators into the aliens. These quantities can be extracted from the direct calculation of the singularities of the OMEs with an insertion of $\Op_{\rm g}^{\text{bare}}$ ($\Op_{\rm q}^{\text{bare}}$). For instance, the coupling $\eta^{\text{bare}}(N)$, which is associated to an operator with a two-point vertex, was determined in \cite{Dixon:1974ss,Hamberg:1991qt} from the renormalization of the OMEs of $\Op_{\rm g}$ with two external ghosts and it was found to be \footnote{Note that the expression for $\eta$ in \cite{Hamberg:1991qt} has an additional factor of 2. This is a consequence of the chosen conventions for dimensional regularization. In particular, we use $D=4-2\eps$ while \cite{Hamberg:1991qt} employs $D=4+\eps$.}
\begin{equation}
   \eta^{\text{bare}}(N)=Z_{\mathrm{g}\,c} = -\frac{a_s}{\epsilon}\frac{{C_A}}{N(N-1)}+{\cal{O}}(a_s^2),
\end{equation}
where $C_A$ is the quadratic Casimir in the adjoint representation and 
$a_s=\alpha_s/(4\pi)=g_s^2/(4\pi)^2$. 
The value of $\eta$ was determined at two loops in \cite{Hamberg:1991qt} and at three loops in \cite{Gehrmann:2023ksf}. Throughout this paper we will mainly be interested in the one-loop alien couplings. As such, it will be convenient to select just the $N$-dependent part of the one-loop result of $\eta^{\text{bare}}(N)$, which in the following we simply denote by $\eta(N)$, i.e.
\begin{equation}
    \eta(N) = \frac{-1}{N(N-1)}.
\end{equation}
The couplings of the operators featuring multiple fields, e.g., the couplings $\kappa_{ij}$ multiply {\textit{at least}} three fields, are determined by renormalizing OMEs with the corresponding external fields. Recently, a method to compute the counterterms of the OMEs with insertions of the gauge-invariant operators as a function of the spin $N$ was put forward in ref.~\cite{Gehrmann:2023ksf}. The result of that paper can be used to extract the coefficients $\kappa^{\text{bare}}_{i\,j}$ up to ${\cal{O}}(a_s^2)$ and those of $\kappa^{(p);\text{bare}}_{ijk}$, for $p=1,2$, at ${\cal{O}}(a_s)$, finding agreement with the low-$N$ values reported in \cite{Falcioni:2022fdm}. In addition, the calculation of the five-point counterterms at ${\cal{O}}(a_s)$, which can be used to determine $\kappa^{(p)}_{ijkl}$, for $p=1,2$, has been announced recently \cite{Gehrmann:2024ggw}. 

In this paper, we would like to determine the renormalization constants $Z_{\rm gi}$ by solving the constraints on the couplings $\mathcal{C}_{\rm i}^{\text{bare}}$, which are imposed by the symmetries of Eq.~(\ref{eq:RenLagr}). The latter is the Lagrangian in Eq.~(\ref{eq:LagrNONREN}) evaluated with bare fields and couplings constants. Therefore the two Lagrangians share the same symmetry properties, with the obvious substitutions. 
For simplicity, in the rest of this paper we drop the superscript `bare', wherever it does not create any ambiguity. 

\paragraph{Independent operators and couplings}
The symmetry constraints on the couplings in Eq.~(\ref{eq:RenLagr}) have been derived in ref.~\cite{Falcioni:2022fdm}. Without repeating the derivation of that paper, we distinguish three types of relations, which follow from the way we have constructed the operators at the beginning of this section.

First of all, the couplings introduced in the EOM operators, see Eqs.~(\ref{eq:eomI})-(\ref{eq:eomIV}) and (\ref{def:Cij})-(\ref{def:Cijkl}), are chosen to inherit the properties of the colour structures they multiply. For example, because of the anti-symmetry of the structure constants, we take
\begin{equation}
    \kappa_{ij}=-\kappa_{ji}.
\end{equation}
This implies, e.g., that at spin $N=4$, where $i,j=0,1$, there is only one independent coupling, e.g., $\kappa_{0\,1}$.

The second type of constraints regards the couplings that enter the ghost operators, Eqs.~(\ref{eq:ghost-LO})-(\ref{eq:ghost-NNNLO}). Because these operators were constructed directly from the EOM ones using gBRST, the $\eta$ couplings are connected to the $\kappa$ ones. The relevant identities have been listed in Eqs.~(\ref{def:etaij})-(\ref{def:eta2bijkl}). 

Finally, we impose the invariance of Eq.~(\ref{eq:RenLagr}) under the generalized transformations of anti-BRST type \cite{Curci:1976bt,Ojima:1980da,Baulieu:1981sb}, which stem from Eq.~(\ref{eq:gengauge}), by replacing the transformation parameter $\omega^a$ with the anti-ghost field $\bar{c}^a$. This implies the following condition on the ghost operator $\Op_c^{(N)}$ defined in Eq.~(\ref{eq:ghostFULL})
\begin{equation}
    O_{\rm c}^{(N)}(A^a_\mu,c^a,\bar{c}^a) = O_{\rm c}^{(N)}(A^a_\mu,\bar{c}^a,c^a),
\end{equation}
which translates into a set of constraints on the couplings in Eqs.~(\ref{eq:ghost-LO})-(\ref{eq:ghost-NNNLO}) and, in turn, on those of the EOM operators. Taking the example of $N=4$, the anti-BRST relation imposes $\kappa_{01}=2\eta(4)$, thus reducing the number of independent couplings even further \cite{Falcioni:2022fdm}.

It is highly non-trivial to find all-$N$ solutions for all the constraints. In refs.~\cite{Falcioni:2022fdm,Falcioni:2024xyt}, they were solved only for fixed values of $N$, in order to fix bases of independent alien operators up to $N=20$. In the following sections, we solve the relations with exact $N$ dependence. This is done by setting up an ansatz for the function space that enters to leading order in $a_s$. The construction of this ansatz is primarily based on constraints from (anti-)gBRST. We will see below that the latter allow one to bootstrap the functional form of higher-order couplings from that of the lower-order ones. The determination of the unknown parameters in the ansatz is then performed by using the full set of colour, gBRST and anti-gBRST relations. As will become clear below, this allows one to fix most, but not all, free parameters. The few that remain then need to be determined from the explicit renormalization of a limited number of fixed-$N$ operator matrix elements. This is particularly important for finding any overall $N$-dependent function.

\section{Identities among the alien couplings}
\label{sec:identities}

In this section we will discuss in detail the identities between the couplings coming from the (anti)-gBRST relations. In particular, we will show that they allow one to restrict the function space of the couplings and hence constrain their generic $N$-dependence.

\subsection{Class II couplings}
\label{sec:nlo}
The class II operators are defined in terms of two couplings, $\kappa_{ij}$ and $\eta_{ij}$, which obey the following relations
\begin{align}
     &\kappa_{ij}+\kappa_{ji}=0, &[\text{anti-symmetry of }f]\label{eq:kappaij-asym}\\
    &\eta_{ij}=2\kappa_{ij}+\eta(N)\binom{i+j+1}{i}\label{eq:etaij-brst}, &[\text{gBRST}]\\
    &\eta_{ij}+\sum_{s=0}^{i}(-1)^{s+j}\binom{s+j}{j}\eta_{(i-s)(j+s)} = 0\label{eq:eta-abrst}. &[\text{anti-gBRST}]
\end{align}
Note that one can generate an equation for the ghost coupling alone by combining the anti-symmetry of $\kappa_{ij}$, Eq.~(\ref{eq:kappaij-asym}), with the gBRST relation, Eq.~(\ref{eq:etaij-brst}),
\begin{equation}
\label{eq:LO-asym+gBRST}
    \eta_{ij}+\eta_{ji} = \eta(N)\Bigg[\binom{i+j+1}{i}+\binom{i+j+1}{j}\Bigg].
\end{equation}
The one-loop value of this coupling was first computed in \cite{Dixon:1974ss}
and later corrected in  \cite{Hamberg:1991qt}.
In our conventions it reads \footnote{Note that there are typos in the corresponding expression in \cite{Falcioni:2022fdm}. In particular, the right-hand side of Eq.~(4.38) in \cite{Falcioni:2022fdm} should be replaced by the right-hand side of Eq.~(\ref{eq:etaij}) here.
}
\begin{equation}
\label{eq:etaij}
    \eta_{ij}  =-\frac{\eta(N)}{4}\Bigg[(-1)^j-3\binom{N-2}{i+1}-\binom{N-2}{i}\Bigg]
\end{equation}
which implies
\begin{equation}
    \kappa_{ij} = -\frac{\eta(N)}{8}\left[(-1)^{j}+3\binom{N-2}{i}-3\binom{N-2}{i+1}\right].
\end{equation}
The power of the relations described above is that they can be used to gain valuable information about the structure of the couplings at arbitrary $N$. For example, one can use Eq.~(\ref{eq:LO-asym+gBRST}) to write down an ansatz for $\eta_{ij}$ of the form
\begin{equation}
    \eta_{ij} = \eta(N) \Big[ c_1\binom{i+j+1}{i}+c_2\binom{i+j+1}{j} \Big].
\end{equation}
Here $c_1$ and $c_2$ are constants to be determined.  We assume here that the dependence on $\eta(N)$ is factorized at leading order, 
as suggested by Eq.~(\ref{eq:LO-asym+gBRST}) and observed in Eq.~(\ref{eq:etaij}). 

This ansatz can then be substituted in the anti-gBRST consistency relation, Eq.~(\ref{eq:eta-abrst}), yielding
\begin{equation}
\label{eq:agbrst-ansatz}
    \eta_{ij}+\sum_{s=0}^{i}(-1)^{s+j}\binom{s+j}{j}\eta_{(i-s)(j+s)} =c_1\,\eta(N) \left[(-1)^{j}+\binom{i+j+1}{i} \right]
\end{equation}
for even values of $N$. Hence we find a consistent solution when $c_1=0$, while $c_2$ is unconstrained. Assuming furthermore that $\kappa_{ij}$ lives in the same function space as $\eta_{ij}$, which is motivated by their close relationship due to the generalized BRST symmetry, we posit
\begin{equation}
    \kappa_{ij} = \eta(N) \left[ b_1\binom{i+j+1}{i}+b_2\binom{i+j+1}{j} \right].
\end{equation}
Then, requiring the anti-symmetry relation in Eq.~(\ref{eq:kappaij-asym}) and the gBRST one in Eq.~(\ref{eq:etaij-brst}) to hold actually produces a unique solution for both couplings, which reads
\begin{align}
\label{eq:wrongetaij}
    &\eta_{ij} = \eta(N)\binom{N-2}{j},\\
    &\kappa_{ij} = \frac{\eta(N)}{2}\left[\binom{N-2}{j}-\binom{N-2}{i}\right].
\end{align}
Note that we used $i+j=N-3$. Next, we compare with some fixed-$N$ evaluation to check our result. While it is obvious from the actual solution in Eq.~(\ref{eq:etaij}) that the result in Eq.~(\ref{eq:wrongetaij}) is in fact incorrect, there is numerical agreement between both for $N=4$. So, to determine that Eq.~(\ref{eq:wrongetaij}) does not represent the physical solution, we actually need to evaluate one more moment, say $N=6$, for which the disagreement does become obvious. Hence, we now need to extend the ansatz, for which we can use the anti-gBRST relation. In particular, note that the right-hand side of Eq.~(\ref{eq:agbrst-ansatz}) suggests the inclusion of a term proportional to $(-1)^{j}$ to the ansatz,
\begin{equation}
     \eta_{ij} =\eta(N) \left[ c_1\binom{i+j+1}{i}+c_2\binom{i+j+1}{j}+c_3(-1)^j\right].
\end{equation}
The anti-gBRST relation now becomes
\begin{equation}
    \eta_{ij}+\sum_{s=0}^{i}(-1)^{s+j}\binom{s+j}{j}\eta_{(i-s)(j+s)} = \eta(N) (c_1+c_3)\left[(-1)^j+\binom{i+j+1}{i}\right]
\end{equation}
such that $c_3=-c_1$ is a consistent solution. If we now impose also Eq.~(\ref{eq:LO-asym+gBRST}) then we obtain the relation $c_1+c_2=1$, leaving just one free parameter unconstrained. 
Hence
\begin{equation}
    \eta_{ij} =\eta(N) \left[ c_1\left[\binom{i+j+1}{i}-(-1)^j\right]+(1-c_1)\binom{i+j+1}{j} \right].
\end{equation}

It should be noted that, if an ansatz is generated using (anti-)gBRST relations, one is in principle free to add non-zero functions that live in the kernel of these relations. For example, if one adds a term of the form
\begin{equation}
\label{eq:KN}
    -\frac{f(N)}{4}\left((-1)^j + \binom{N - 2}{i + 1} - \binom{N - 2}{i}\right)
\end{equation}
to Eq.~(\ref{eq:etaij}), the corresponding expression for $\eta_{ij}$ still obeys the constraints. Here $f(N)$ represents an arbitrary function of $N$, with the actual solution being recovered by setting $f(N)=0$ for even values of $N$. In particular, substituting Eq.~(\ref{eq:KN}) in the constraint coming from anti-symmetry and gBRST, cf.~Eq.~(\ref{eq:LO-asym+gBRST}), one finds
\begin{equation}
    \left[(-1)^i+(-1)^j \right]f(N)=0.
\end{equation}
The left-hand side of this expression always vanishes for all physical (even) values of $N$, independent of the functional form of $f(N)$. In general, the exclusion of this type of function can only be confirmed by comparison with fixed-$N$ computations.

An important consequence is that now we have recovered the full function space of the actual solution, Eq.~(\ref{eq:etaij}), using only the symmetry relations of the couplings. More generally, note that Eq.~(\ref{eq:eta-abrst}) is an example of a \textit{conjugation} relation, in the sense that a second application of the sum leads to
\begin{equation}
    \sum_{t=0}^{i}(-1)^{t+j}\binom{t+j}{j}\eta_{(i-t)(j+t)} = -\sum_{t=0}^{i}(-1)^{t+j}\binom{t+j}{j}\sum_{s=0}^{i-t}(-1)^{s+j+t}\binom{s+j+t}{j+t}\eta_{(i-t-s)(j+t+s)}
\end{equation}
and hence
\begin{equation}
    \eta_{ij} = \sum_{t=0}^{i}\binom{t+j}{j}\sum_{s=0}^{i-t}(-1)^{s}\binom{s+j+t}{j+t}\eta_{(i-t-s)(j+t+s)}.
\end{equation}
The latter identity is actually \textit{always} true for \textit{any} discrete two-variable function $\eta_{ij}$. This type of conjugation relation has already been encountered in the computation of the anomalous dimensions of twist-two operators in non-forward kinematics, see e.g.~\cite{Moch:2021cdq,VanThurenhout:2023gmo}, and holds great predictive power. In particular, it provides valuable information about the function space of the object at hand. To take full advantage of such relations, one needs to be able to evaluate them analytically. This is possible by using principles of symbolic summation, in particular by application of the creative telescoping algorithm \cite{Zeilberger1991}. The latter is a generalization of classical telescoping and attempts to evaluate the sum of interest by rewriting it as a recursion relation using Gosper's algorithm \cite{Gosper1978}. The closed-form expression of the sum then corresponds to the linear combination of the solutions of the recursion that has the same initial values as the sum. This methodology is neatly implemented in the {\tt Mathematica} package {\tt Sigma} \cite{Schneider2004,Schneider2007}. For the class III and IV couplings to be described below we will also encounter identities involving multiple sums, for which the package {\tt EvaluateMultiSums} \cite{Schneider:2013uan,Schneider:2013zna} can be used.

\subsection{Class III couplings}
\label{sec:n2lo}

\subsubsection{$\kappa_{ijk}^{(1)}$ and $\eta_{ijk}^{(1)}$}
\label{sec:kappa1-ijk}
The couplings $\eta_{ijk}^{(1)}$ and $\kappa_{ijk}^{(1)}$ can be thought of as direct generalizations of $\eta_{ij}$ and $\kappa_{ij}$ in the class II operators. They obey the following relations
\begin{align}
     &\kappa_{ijk}^{(1)}+\kappa_{ikj}^{(1)}=0, &[\text{anti-symmetry of }f]\label{eq:kappa1-antisym}\\
    &\kappa_{ijk}^{(1)}+\kappa_{jki}^{(1)}+\kappa_{kij}^{(1)} = 0, &[\text{Jacobi identity}]\label{eq:kappa1-jacobi}\\
    &\eta_{ijk}^{(1)}=2\kappa_{i(j+k+1)}\binom{j+k+1}{j}+2[\kappa_{ijk}^{(1)}+\kappa_{kji}^{(1)}]\label{eq:kappa1-gBRST}, &[\text{gBRST}]\\
    &\eta_{ijk}^{(1)}=\sum_{m=0}^{i}\sum_{n=0}^{j}\frac{(m+n+k)!}{m!n!k!}(-1)^{m+n+k}\eta_{(j-n)(i-m)(k+m+n)}^{(1)}\label{eq:eta1-agBRST}. &[\text{anti-gBRST}]
\end{align}
Note that now the indices are constrained as $i+j+k=N-4$. As before, one can combine the relations of the EOM coupling with the gBRST relation to connect $\eta_{ijk}^{(1)}$ to $\kappa_{ijk}^{(1)}$. In particular we find
\begin{equation}
    \eta_{ijk}^{(1)}+\eta_{ikj}^{(1)} = 2\kappa_{i(j+k+1)}\binom{j+k+2}{j+1}+2[\kappa_{kji}^{(1)}+\kappa_{jki}^{(1)}]
\end{equation}
when combining the anti-symmetry property of $\kappa_{ijk}^{(1)}$ with the gBRST identity. Similarly the combination of the Jacobi identity with gBRST leads to
\begin{dmath}
\label{eq:eta1-Jacobi+gBRST}
    \eta_{ijk}^{(1)}+\eta_{kij}^{(1)}+\eta_{jki}^{(1)} = 2\kappa_{i(j+k+1)}\binom{j+k+1}{j}+2\kappa_{k(i+j+1)}\binom{i+j+1}{i}+2\kappa_{j(i+k+1)}\binom{i+k+1}{k}.
\end{dmath}
The latter identity relates the class III coupling $\eta_{ijk}^{(1)}$, which is ${\cal O}(g_s^{2})$, 
to the class II coupling $\kappa_{ij}$ of ${\cal O}(g_s)$, i.e.\ at one order lower in perturbation theory. 
As such, we can use it to determine the function space of $\eta_{ijk}^{(1)}$. Taking into account all independent permutations of $i$, $j$ and $k$, we find that this function space is 18-dimensional
\begin{align}
\label{eq:funspace}
    \Bigg\{&(-1)^{i+j}\binom{i+j+1}{i},\binom{N-2}{k+1}\binom{i+j+1}{i},\binom{N-2}{k}\binom{i+j+1}{i},(-1)^{j+k}\binom{j+k+1}{j},\nonumber\\&\binom{N-2}{i+1}\binom{j+k+1}{j},\binom{N-2}{i}\binom{j+k+1}{j},(-1)^{i+k}\binom{i+k+1}{k},\binom{N-2}{j+1}\binom{i+k+1}{k},\nonumber\\&\binom{N-2}{j}\binom{i+k+1}{k} + \text{ independent permutations of $i$, $j$ and $k$}\Bigg\}.
\end{align}
Furthermore, due to the close relationship between $\eta_{ijk}^{(1)}$ and $\kappa_{ijk}^{(1)}$, we assume that the functional form of the latter is constructed from the same functions.  
Hence in total we have 36 free parameters. Using the relations described above, cf.~Eqs.~(\ref{eq:kappa1-antisym})-(\ref{eq:eta1-agBRST}), we are able to fix 34 of these. The final two free parameters are then determined using the one-loop results $\kappa_{110}^{(1)}=0$ and $\kappa_{121}^{(1)}=13\:C_A/336$, which follow from the explicit operator renormalization for $N=6$ and $N=8$ respectively. Our final result for $\kappa_{ijk}^{(1)}$ then becomes
\begin{dmath}
\label{eq:kappa1-final}
    \kappa_{ijk}^{(1)} = \frac{\eta(N)}{48}\Bigg\{2(-1)^{i + j}\binom{i+j+1}{ i}+ (-1)^{i + k}\binom{i+k+1}{ k} + 3(-1)^{j + k+1}\binom{j+k+1}{j} +\binom{i+k+1}{ i}\Bigg[ 2(-1)^{i + k+1} + 5\binom{N-1}{j+1}\Bigg] + \binom{j+k+1}{k}\Bigg[3(-1)^{j + k} - 
   10\binom{N-2}{i} + 4\binom{N-2}{i+1}\Bigg] + 
   \binom{i+j+1}{j}\Bigg[(-1)^{i+j+1} + 5\binom{N-2}{k} - 9\binom{N-2}{k+1}\Bigg] \Bigg\}.
\end{dmath}
To verify this expression, agreement with explicitly computed fixed-$N$ values has been established up to $N=20$. The necessary direct computations at fixed values of $N$ of Feynman diagrams for the OMEs 
with (physical or alien) spin-$N$ twist-two operators $\Op^{(N)}$ inserted 
in Green's functions with off-shell quarks, gluons or ghosts are performed with the setup used and described in \cite{Moch:2017uml,Falcioni:2023luc,Falcioni:2023vqq,Moch:2023tdj,Falcioni:2024xyt} for the computation of moments of four-loop QCD splitting functions. 
In particular, the {\tt Forcer} package~\cite{Ruijl:2017cxj}, written in {\tt Form}~\cite{Vermaseren:2000nd,Kuipers:2012rf,Ruijl:2017dtg}, is used for the parametric reductions of the two-point functions up to four loops for fixed even integer values of $N$. Substituting our result for $\kappa_{ijk}^{(1)}$ into the gBRST relation, Eq.~(\ref{eq:kappa1-gBRST}), allows one to also reconstruct the full $N$-dependence of $\eta_{ijk}^{(1)}$
\begin{dmath}
\label{eq:eta1-final}
    \eta_{ijk}^{(1)} = -\frac{\eta(N)}{24}\Bigg\{5(-1)^{i+j+1}\binom{i+j+1}{i}+(-1)^{i+k}\binom{i+k+1}{k}+2(-1)^{j+k+1}\binom{j+k+1}{j}+\binom{i+k+1}{i}\Bigg[(-1)^{i+k}+4\binom{N-2}{j+1}\Bigg]+\binom{j+k+1}{k}\Bigg[5(-1)^{j+k+1}-3\binom{N-2}{i}+\binom{N-2}{i+1}\Bigg]+\binom{i+j+1}{j}\Bigg[4(-1)^{i+j}-15\binom{N-2}{k}-5\binom{N-2}{k+1}\Bigg]\Bigg\}.
\end{dmath}
We have verified that Eqs.~(\ref{eq:kappa1-final}) and (\ref{eq:eta1-final}) are in agreement with the results of ref.~\cite{Gehrmann:2023ksf}, as explained in Sec.~\ref{sec:feynman} below.
\subsubsection{$\kappa_{ijk}^{(2)}$ and $\eta_{ijk}^{(2)}$}
\label{sec:kappa2-ijk}
The next alien couplings we consider are $\kappa_{ijk}^{(2)}$ and $\eta_{ijk}^{(2)}$, which obey the following relations
\begin{align}
    &\kappa_{ijk}^{(2)}=\kappa_{jik}^{(2)}=\kappa_{ikj}^{(2)}=\kappa_{kji}^{(2)}=\kappa_{jki}^{(2)}=\kappa_{kij}^{(2)}, \label{eq:kappa2-sym} &[\text{symmetry of $d_4$}]\\
    &\eta_{ijk}^{(2)} = 3\kappa_{ijk}^{(2)} \label{eq:kappa2-gBRST}, &[\text{gBRST}] \\
    &\eta_{ijk}^{(2)} = \sum_{m=0}^{i}\sum_{n=0}^{j}(-1)^{m+n+k}\frac{(m+n+k)!}{m!n!k!}\eta_{(i-m)(j-n)(m+n+k)}^{(2)}. \label{eq:eta2-agBRST}&[\text{anti-gBRST}]
\end{align}
As the anti-gBRST equation has a similar form as the one for $\eta_{ijk}^{(1)}$, cf.~Eq.~(\ref{eq:eta1-agBRST}), we assume the function space for $\eta_{ijk}^{(2)}$ and $\kappa_{ijk}^{(2)}$ to be the same as above, cf.~Eq.~(\ref{eq:funspace}). Imposing Eqs.~(\ref{eq:kappa2-sym})-(\ref{eq:eta2-agBRST}) then allows one to fix all but one of the unknowns. Hence we find expressions $\eta_{ijk}^{(2)}$ and $\kappa_{ijk}^{(2)}$ with only one (overall) free parameter
\begin{align}
    &\kappa_{ijk}^{(2)} = c\Bigg\{(-1)^{i+j}\binom{i+j+2}{i+1}+(-1)^{i+k}\binom{i+k+2}{i+1}+\binom{j+k+2}{j+1}\Bigg[(-1)^{j+k}+\binom{N-1}{i+1}\Bigg]\Bigg\},\label{eq:kappa2-finA}\\
    &\eta_{ijk}^{(2)} = 3\kappa_{ijk}^{(2)}.\label{eq:eta2-finA}
\end{align}
Note that the $c$ parameter can a priori be some $N$-dependent function. A computation of the OMEs at a few fixed values of $N$ with the procedure outlined in Sec.~\ref{sec:kappa1-ijk} for the renormalization of the respective operators fixes $c=1/N/(N-1)$ such that
\begin{align}
    &\kappa_{ijk}^{(2)} = \frac{1}{N(N-1)}\Bigg\{(-1)^{i+j}\binom{i+j+2}{i+1}+(-1)^{i+k}\binom{i+k+2}{i+1}+\binom{j+k+2}{j+1}\Bigg[(-1)^{j+k}+\binom{N-1}{i+1}\Bigg]\Bigg\},\label{eq:kappa2-fin}\\
    &\eta_{ijk}^{(2)} = 3\kappa_{ijk}^{(2)}.\label{eq:eta2-fin}
\end{align}
Noting that
\begin{equation}
    \frac{1}{N(N-1)} = -\eta(N)
\end{equation}
we see that also for these two couplings $\eta(N)$ factorizes, which is not expected a priori from the constraints.

\subsubsection{$\kappa_{ijk}^{(3)}$ and $\eta_{ijk}^{(3)}$}
\label{sec:kappa3-ijk}
The last set of couplings in the class III alien operators, $\kappa_{ijk}^{(3)}$ and $\eta_{ijk}^{(3)}$, obey the following relations
\begin{align}
    &\kappa_{ijk}^{(3)}=\kappa_{ikj}^{(3)}, \label{eq:kappa3-sym} &[\text{symmetry}] \\
    &\kappa_{ijk}^{(3)}+\kappa_{kij}^{(3)}+\kappa_{jki}^{(3)}=0, \label{eq:kappa3-jacobi} &[\text{generalized Jacobi identity}] \\
    &\eta_{ijk}^{(3)} = 2(\kappa_{ijk}^{(3)}-\kappa_{kji}^{(3)}),\label{eq:kappa3-gbrst} &[\text{gBRST}]\\
    &\eta_{ijk}^{(3)} = \sum_{m=0}^{i}\sum_{n=0}^{j}(-1)^{m+n+k}\frac{(m+n+k)!}{m!n!k!}\eta_{(j-n)(i-m)(m+n+k)}^{(3)}. \label{eq:eta3-agBRST} &[\text{anti-gBRST}]
\end{align}
As before, we suggest the same function space as for $\kappa_{ijk}^{(1)}$ and $\eta_{ijk}^{(1)}$, cf.~Eq.~(\ref{eq:funspace}). The above relations then only leave two parameters unfixed, such that we have
\begin{dmath}
\label{eq:kappa3-fin}
    \kappa_{ijk}^{(3)} = c_1 (-1)^{i+j} \binom{i+j+1}{i}+c_2 (-1)^{i+k} 
\binom{i+k+1}{k}+\binom{j+k+1}{j}\Bigg[(c_1+c_2) (-1)^{j+k+1} 
+(c_1+c_2) (-1)^{j+k+1} +c_1 \binom{N-2}{i+1}\Bigg]+\binom{i+k+1}{i}\Bigg[c_1 (-1)^{i+k}+(2 c_1+c_2) 
\binom{N-2}{j}+c_2 \binom{N-2}{j+1}\Bigg]+\binom{i+j+1}{j}\Bigg[c_2 (-1)^{i+j}-(2 c_1+c_2) \binom{N-2}{k}-(c_1+c_2)  \binom{N-2}{k+1}\Bigg]
\end{dmath}
with $c_1, c_2$ to be determined. We emphasize that, as before, these could be $N$-dependent functions \footnote{In this case we expect $c_1\sim c_2\sim \eta(N)$.}. The corresponding expression for $\eta_{ijk}^{(3)}$ depends on the same parameters through the gBRST relation, cf.~Eq.~(\ref{eq:kappa3-gbrst}).
Since the couplings $\kappa_{ijk}^{(3)}$ and $\eta_{ijk}^{(3)}$ do not appear through operator mixing in the renormalization of physical OMEs up to four loops, we leave the two free parameters $c_1, c_2$ in Eq.~(\ref{eq:kappa3-fin}) undetermined, for the time being. 
We will address this issue again when extending the computation of low-$N$ non-singlet anomalous dimensions at five loops~\cite{Herzog:2018kwj} to the flavor-singlet sector.

\subsection{Class IV couplings}
\label{sec:n3lo}

\subsubsection{$\kappa_{ijkl}^{(1)}$ and $\eta_{ijkl}^{(1)}$}
We have the following set of relations
\begin{align}
    &\kappa_{ijkl}^{(1)}+\kappa_{ijlk}^{(1)} = 0,\label{eq:kappa1-n3lo-asym}&[\text{anti-symmetry}] \\
    &\kappa_{ijkl}^{(1)}+\kappa_{iklj}^{(1)}+\kappa_{iljk}^{(1)} = 0, \label{eq:kappa1-n3lo-jacobi} &[\text{Jacobi}] \\
    &\kappa_{ijkl}^{(1)}+\kappa_{jilk}^{(1)}+\kappa_{lkji}^{(1)}+\kappa_{klij}^{(1)} = 0,\label{eq:kappa1-n3lo-djacobi} &[\text{double Jacobi}] \\
    &\eta_{ijkl}^{(1)} = 2[\kappa_{ij(l+k+1)}^{(1)}+\kappa_{(l+k+1)ji}^{(1)}]\binom{l+k+1}{k}+2[\kappa_{ijkl}^{(1)}+\kappa_{ilkj}^{(1)}+\kappa_{likj}^{(1)}+\kappa_{lkij}^{(1)}],\label{eq:kappa1-n3lo-gbrst} & [\text{gBRST}] \\
    &\eta_{ijkl}^{(1)} = -\sum_{s_1=0}^{i}\sum_{s_2=0}^{j}\sum_{s_3=0}^{k}\frac{(s_1+s_2+s_3+l)!}{s_1!s_2!s_3!l!}(-1)^{s_1+s_2+s_3+l}\eta_{(k-s_3)(j-s_2)(i-s_1)(s_1+s_2+s_3+l)}^{(1)}\label{eq:eta1-n3lo-agbrst} &  [\text{anti-gBRST}] 
\end{align}
with now $i+j+k+l=N-5$.
Combining the double Jacobi identity, Eq.~(\ref{eq:kappa1-n3lo-djacobi}), with the gBRST one, Eq.~(\ref{eq:eta3-agBRST}), allows one to write $\eta_{ijkl}^{(1)}$ in terms of $\kappa_{ijk}^{(1)}$ appearing already in the class III operators at one order in perturbation theory lower,
\begin{dmath}
    \eta_{ijkl}^{(1)} + \eta_{jilk}^{(1)} + \eta_{lkji}^{(1)} + \eta_{klij}^{(1)} = 2[\kappa_{ij(k+l+1)}^{(1)}+\kappa_{(k+l+1)ji}^{(1)}]\binom{k+l+1}{k} + 2[\kappa_{ji(k+l+1)}^{(1)}+\kappa_{(k+l+1)ij}^{(1)}]\binom{k+l+1}{l}  + 2[\kappa_{lk(i+j+1)}^{(1)}+\kappa_{(i+j+1)kl}^{(1)}]\binom{i+j+1}{j}  + 2[\kappa_{kl(i+j+1)}^{(1)}+\kappa_{(i+j+1)lk}^{(1)}]\binom{i+j+1}{i}.
\end{dmath}
As such, we can use the expression we have computed for $\kappa_{ijk}^{(1)}$, cf.~Eq.~(\ref{eq:kappa1-final}), to determine the function space of $\eta_{ijkl}^{(1)}$. Taking into account all the independent permutations of the indices $i, k, j$ and $l$ this space is now 264-dimensional. Assuming that the functional form of $\kappa_{ijkl}^{(1)}$ is similar to the one of $\eta_{ijkl}^{(1)}$ then implies that in total we now have 528 parameters to fix. However, after implementing all of the above relations, only 8 remain in the end. The latter can again be fixed from the explicit renormalization of a few fixed-$N$ matrix elements. More specifically we extracted them by performing a small momentum expansion around the limit $p_3,p_4,p_5 \to 0$ of the OME 
\begin{equation}
\label{eq:OMEccbggg}
\langle O_{\rm g}^{(N)}; \bar c(p_1) c(p_2) g(p_3) g(p_4) g(p_5)\rangle \,.
\end{equation}
This expansion is achieved on a diagram-by-diagram basis using the expansion-by-subgraph method \cite{Smirnov:1990rz,Smirnov:1994tg,Smirnov:2002pj} 
to second order at $N=10$ and to third order at $N=12$. By expanding sequentially in the external gluon momenta $p_3,p_4$ and $p_5$  the integrals are reduced to simple one-scale propagator integrals. We have implemented the expansion-by-subgraph in {\tt Maple} \cite{maple} and then subsequently evaluated the expressions in {\tt Form}. This methodology was also used to cross-check the expressions for $\kappa_{ij}^{(1)}$ and $\kappa_{ijk}^{(r=1,2)}$ up to $N=20$. At one loop the poles of the OME, Eq.~(\ref{eq:OMEccbggg}), are generated purely by the ghost alien operator $O_c^{(N),IV}$ allowing for a clean extraction of $\eta_{ijkl}^{(1)}$ renormalization constants, from which the  $\kappa_{ijkl}^{(1)}$ values can be obtained. In particular, in order to determine the remaining constants in the all-$N$ ansatz for $\kappa_{ijkl}^{(1)}$, we use
\begin{align}
    \kappa^{(1)}_{0210}=-\frac{1}{128}C_A,\quad
    \kappa^{(1)}_{0050}=\frac{109}{1440}C_A,\quad
    \kappa^{(1)}_{0104}=-\frac{935}{6912}C_A,\quad
    \kappa^{(1)}_{1006}=-\frac{2537}{16896}C_A.
\end{align}
We then find\newpage
{\footnotesize\begin{dmath}
\kappa_{ijkl}^{(1)}=-\frac{\eta(N)}{384}\Bigg\{\Bigg[6(-1)^{j+k}\binom{i+l+1}{i}-3(-1)^{j+k}\binom{i+l+1}{l}+7(-1)^{j+k+l}\binom{j+k+l+2}{l}+7(-1)^{j+k+l}\binom{j+k+l+2}{j+k+1}-27\binom{N-1}{i+1}\binom{j+k+l+2}{j+k+1}+2\binom{i+j+k+2}{j+k+1}\Bigg[2(-1)^{i+j+k}+9\binom{N-2}{l}\Bigg]-2\binom{i+j+k+2}{i}\Bigg[4(-1)^{i+j+k}+15\binom{N-2}{l+1}\Bigg]\Bigg]\binom{j+k+1}{j}-\Bigg[5(-1)^{j+k}\binom{i+l+1}{i}-4(-1)^{j+k}\binom{i+l+1}{l}-14(-1)^{j+k+l}\binom{j+k+l+2}{l}+7(-1)^{j+k+l}\binom{j+k+l+2}{j+k+1}+54\binom{N-2}{i}\binom{j+k+l+2}{j+k+1}+\binom{i+j+k+2}{j+k+1}\Bigg[-3(-1)^{i+j+k}+4\binom{N-1}{l+1}\Bigg]+\binom{i+j+k+2}{i}\Bigg[3(-1)^{i+j+k}+13\binom{N-1}{l+1}\Bigg]\Bigg]\binom{j+k+1}{k}+(-1)^{i+j+k+1}\binom{i+k+1}{k}\binom{i+j+k+2}{j}-2(-1)^{i+j+k}\binom{i+j+1}{j}\binom{i+j+k+2}{k}-6(-1)^{j+l}\binom{i+k+1}{i}\binom{j+l+1}{j}+3(-1)^{j+l}\binom{i+k+1}{k}\binom{j+l+1}{j}+5(-1)^{j+l}\binom{i+k+1}{i}\binom{j+l+1}{l}-4(-1)^{j+l}\binom{i+k+1}{k}\binom{j+l+1}{l}+30\binom{N-2}{k+1}\binom{j+l+1}{j}\binom{i+j+l+2}{i}-5(-1)^{i+j+l}\binom{j+l+1}{l}\binom{i+j+l+2}{i}+13\binom{N-1}{k+1}\binom{j+l+1}{l}\binom{i+j+l+2}{i}-4(-1)^{i+j+l}\binom{i+l+1}{i}\binom{i+j+l+2}{j}+12\binom{N-2}{k}\binom{i+l+1}{i}\binom{i+j+l+2}{j}+24\binom{N-2}{k+1}\binom{i+l+1}{i}\binom{i+j+l+2}{j}-3(-1)^{i+j+l}\binom{i+l+1}{l}\binom{i+j+l+2}{j}+49\binom{N-1}{k+1}\binom{i+l+1}{l}\binom{i+j+l+2}{j}-30\binom{N-1}{k+1}\binom{i+j+1}{i}\binom{i+j+l+2}{l}+2(-1)^{i+j+l}\binom{i+j+1}{j}\binom{i+j+l+2}{l}-60\binom{N-2}{k+1}\binom{i+j+1}{j}\binom{i+j+l+2}{l}+8(-1)^{i+j+l}\binom{i+l+1}{i}\binom{i+j+l+2}{i+l+1}-6\binom{N-2}{k+1}\binom{i+l+1}{i}\binom{i+j+l+2}{i+l+1}-3(-1)^{i+j+l}\binom{i+l+1}{l}\binom{i+j+l+2}{i+l+1}+71\binom{N-1}{k+1}\binom{i+l+1}{l}\binom{i+j+l+2}{i+l+1}-11(-1)^{k+l}\binom{i+j+1}{i}\binom{k+l+1}{k}+7(-1)^{k+l}\binom{i+j+1}{j}\binom{k+l+1}{k}+11(-1)^{k+l}\binom{i+j+1}{i}\binom{k+l+1}{l}-7(-1)^{k+l}\binom{i+j+1}{j}\binom{k+l+1}{l}+60\binom{N-2}{j+1}\binom{k+l+1}{k}\binom{i+k+l+2}{i}-10(-1)^{i+k+l}\binom{k+l+1}{l}\binom{i+k+l+2}{i}+26\binom{N-1}{j+1}\binom{k+l+1}{l}\binom{i+k+l+2}{i}+(-1)^{i+k+l+1}\binom{i+l+1}{i}\binom{i+k+l+2}{k}-4\binom{N-1}{j+1}\binom{i+l+1}{i}\binom{i+k+l+2}{k}-2(-1)^{i+k+l}\binom{i+l+1}{l}\binom{i+k+l+2}{k}-44\binom{N-2}{j}\binom{i+l+1}{l}\binom{i+k+l+2}{k}-26\binom{N-2}{j+1}\binom{i+l+1}{l}\binom{i+k+l+2}{k}-15\binom{N-1}{j+1}\binom{i+k+1}{i}\binom{i+k+l+2}{l}+(-1)^{i+k+l}\binom{i+k+1}{k}\binom{i+k+l+2}{l}-30\binom{N-2}{j+1}\binom{i+k+1}{k}\binom{i+k+l+2}{l}+5(-1)^{i+k+l}\binom{i+l+1}{i}\binom{i+k+l+2}{i+l+1}-10\binom{N-1}{j+1}\binom{i+l+1}{i}\binom{i+k+l+2}{i+l+1}+(-1)^{i+k+l}\binom{i+l+1}{l}\binom{i+k+l+2}{i+l+1}-18\binom{N-2}{j+1}\binom{i+l+1}{l}\binom{i+k+l+2}{i+l+1}-14(-1)^{j+k+l}\binom{k+l+1}{l}\binom{j+k+l+2}{j}-7(-1)^{j+k+l}\binom{j+l+1}{l}\binom{j+k+l+2}{k}\Bigg\}
\end{dmath}}
and
{\footnotesize\begin{dmath}
\eta_{ijkl}^{(1)}=-\frac{\eta(N)}{96}\Bigg\{\Bigg[\Bigg[\Bigg(\binom{j+k+l+2}{l}+5\binom{j+k+l+2}{j+k+1}\Bigg)(-1)^{l+1}+3\binom{i+l+1}{i}+3\binom{i+l+1}{l}\Bigg](-1)^{j+k}-17(-1)^{i+j+k}\binom{i+j+k+2}{i}+\binom{i+j+k+2}{j+k+1}\Bigg[13(-1)^{i+j+k}+54\binom{N-2}{l}\Bigg]\Bigg]\binom{j+k+1}{j}+\Bigg[-3(-1)^{j+k}\binom{i+l+1}{i}-3(-1)^{j+k}\binom{i+l+1}{l}+17(-1)^{j+k+l}\binom{j+k+l+2}{l}+7(-1)^{j+k+l}\binom{j+k+l+2}{j+k+1}+6\binom{N-2}{i}\binom{j+k+l+2}{j+k+1}+\binom{i+j+k+2}{i}\Bigg[(-1)^{i+j+k}+6\binom{N-1}{l+1}\Bigg]+\binom{i+j+k+2}{j+k+1}\Bigg[(-1)^{i+j+k}+6\binom{N-1}{l+1}\Bigg]\Bigg]\binom{j+k+1}{k}-12(-1)^{i+j+k}\binom{i+j+1}{j}\binom{i+j+k+2}{k}+(-1)^{j+l+1}\binom{i+k+1}{i}\binom{j+l+1}{j}+(-1)^{j+l+1}\binom{i+k+1}{k}\binom{j+l+1}{j}+(-1)^{j+l+1}\binom{i+k+1}{i}\binom{j+l+1}{l}+(-1)^{j+l+1}\binom{i+k+1}{k}\binom{j+l+1}{l}-3(-1)^{i+j+l}\binom{i+l+1}{i}\binom{i+j+l+2}{j}-18\binom{N-2}{k}\binom{i+l+1}{i}\binom{i+j+l+2}{j}+18\binom{N-2}{k+1}\binom{i+l+1}{i}\binom{i+j+l+2}{j}-3(-1)^{i+j+l}\binom{i+l+1}{l}\binom{i+j+l+2}{j}+3\binom{N-1}{k+1}\binom{i+l+1}{l}\binom{i+j+l+2}{j}+18\binom{N-1}{k+1}\binom{i+j+1}{i}\binom{i+j+l+2}{l}-6(-1)^{i+j+l}\binom{i+j+1}{j}\binom{i+j+l+2}{l}-30\binom{N-2}{k+1}\binom{i+j+1}{j}\binom{i+j+l+2}{l}+3(-1)^{i+j+l}\binom{i+l+1}{i}\binom{i+j+l+2}{i+l+1}+12\binom{N-2}{k+1}\binom{i+l+1}{i}\binom{i+j+l+2}{i+l+1}+3(-1)^{i+j+l}\binom{i+l+1}{l}\binom{i+j+l+2}{i+l+1}+3\binom{N-1}{k+1}\binom{i+l+1}{l}\binom{i+j+l+2}{i+l+1}+7(-1)^{k+l}\binom{i+j+1}{i}\binom{k+l+1}{k}-5(-1)^{k+l}\binom{i+j+1}{j}\binom{k+l+1}{k}+17(-1)^{k+l}\binom{i+j+1}{i}\binom{k+l+1}{l}-13(-1)^{k+l}\binom{i+j+1}{j}\binom{k+l+1}{l}-18\binom{N-2}{j+1}\binom{k+l+1}{k}\binom{i+k+l+2}{i}-2(-1)^{i+k+l}\binom{k+l+1}{l}\binom{i+k+l+2}{i}-10\binom{N-1}{j+1}\binom{k+l+1}{l}\binom{i+k+l+2}{i}+(-1)^{i+k+l+1}\binom{i+l+1}{i}\binom{i+k+l+2}{k}-5\binom{N-1}{j+1}\binom{i+l+1}{i}\binom{i+k+l+2}{k}+(-1)^{i+k+l+1}\binom{i+l+1}{l}\binom{i+k+l+2}{k}+10\binom{N-2}{j}\binom{i+l+1}{l}\binom{i+k+l+2}{k}-26\binom{N-2}{j+1}\binom{i+l+1}{l}\binom{i+k+l+2}{k}+(-1)^{i+k+l+1}\binom{i+l+1}{i}\binom{i+k+l+2}{i+l+1}-5\binom{N-1}{j+1}\binom{i+l+1}{i}\binom{i+k+l+2}{i+l+1}-3(-1)^{i+k+l}\binom{i+l+1}{l}\binom{i+k+l+2}{i+l+1}-24\binom{N-2}{j+1}\binom{i+l+1}{l}\binom{i+k+l+2}{i+l+1}+4(-1)^{j+k+l}\binom{k+l+1}{l}\binom{j+k+l+2}{j}\Bigg\}.
\end{dmath}}
We have checked the correctness of these expressions by comparing with fixed-$N$ computations up to $N=14$.

\subsubsection{$\kappa_{ijkl}^{(2)}$, $\eta_{ijkl}^{(2a)}$ and $\eta_{ijkl}^{(2b)}$}
\label{sec:kappa2-ijkl}
For this final set of couplings we have the following relations
\begin{align}
    &\kappa_{ijkl}^{(2)}+\kappa_{ijlk}^{(2)} = 0,\label{eq:kappa2-n3lo-asym}&[\text{anti-symmetry}] \\
    &\kappa_{ijkl}^{(2)} = \kappa_{jikl}^{(2)},\label{eq:kappa2-n3lo-sym}&[\text{symmetry of $d_{4}$}] \\
    &\eta_{ijkl}^{(2a)} = 3\kappa_{ij(k+l+1)}^{(2)}\binom{k+l+1}{k}+2\kappa_{ijkl}^{(2)},\label{eq:kappa2-n3lo-gbrsta} & [\text{gBRST (a)}] \\
    &\eta_{ijkl}^{(2b)} = 2\kappa_{lijk}^{(2)},\label{eq:kappa2-n3lo-gbrstb} & [\text{gBRST (b)}] \\
    &\eta_{ijkl}^{(2a)} = -\sum_{s_1=0}^{i}\sum_{s_2=0}^{j}\sum_{s_3=0}^{k}\frac{(s_1+s_2+s_3+l)!}{s_1!s_2!s_3!l!}(-1)^{s_1+s_2+s_3+l}\, \times\nonumber\\& \hspace*{20mm}\times\, \eta_{(i-s_1)(j-s_2)(k-s_3)(s_1+s_2+s_3+l)}^{(2a)},\label{eq:eta2a-n3lo-agbrst} &  [\text{anti-gBRST (a)}]  \\
    &\eta_{ijkl}^{(2b)} = \eta_{ikjl}^{(2a)} - \eta_{ijkl}^{(2a)} + \sum_{s_1=0}^{i}\sum_{s_2=0}^{j}\sum_{s_3=0}^{k}\frac{(s_1+s_2+s_3+l)!}{s_1!s_2!s_3!l!}(-1)^{s_1+s_2+s_3+l}\, \times\nonumber\\& \hspace*{20mm}\times\, \eta_{(i-s_1)(j-s_2)(k-s_3)(s_1+s_2+s_3+l)}^{(2b)}.\label{eq:eta2b-n3lo-agbrst} &  [\text{anti-gBRST (b)}]
\end{align}
Note that Eqs.~(\ref{eq:kappa2-n3lo-asym}) and (\ref{eq:kappa2-n3lo-gbrsta}) can be combined to express $\eta_{ijkl}^{(2a)}$ in terms of the class III coupling $\kappa_{ijk}^{(2)}$ as
\begin{equation}
    \eta_{ijkl}^{(2a)} + \eta_{ijlk}^{(2a)} = 3\kappa_{ij(k+l+1)}^{(2)}\binom{k+l+2}{k+1}.
\end{equation}
Using the expression we derived for $\kappa_{ijk}^{(2)}$, cf.~Eq.~(\ref{eq:kappa2-fin}), this becomes
\begin{dmath}
   \eta_{ijkl}^{(2a)} + \eta_{ijlk}^{(2a)} = 3c\Bigg\{(-1)^{i+j}\binom{i+j+2}{i+1}-(-1)^{i+k+l}\binom{i+k+l+3}{i+1}+\binom{j+k+l+3}{j+1}\Bigg[-(-1)^{j+k+l}+\binom{N-1}{i+1}\Bigg]\Bigg\}\binom{k+l+2}{k+1}
\end{dmath}
with $c$ to be determined. Likewise one can use Eqs.~(\ref{eq:kappa2-n3lo-sym}) and (\ref{eq:kappa2-n3lo-gbrsta}) to write
\begin{equation}
    \eta_{ijkl}^{(2a)} - \eta_{jikl}^{(2a)} = 0.
\end{equation}
To obtain this last identity we used the symmetry property of $\kappa_{ijk}^{(2)}$, cf.~Eq.~(\ref{eq:kappa2-sym}).
The complete solution of the gBRST constraints in Eqs.~(\ref{eq:kappa2-n3lo-asym})-(\ref{eq:eta2a-n3lo-agbrst}) proceeds in complete analogy to the previous cases.
However, similar to the class III couplings in Sec.~\ref{sec:kappa3-ijk}, also 
$\kappa_{ijkl}^{(2)}$, $\eta_{ijkl}^{(2a)}$ and $\eta_{ijkl}^{(2b)}$ 
do not enter in the operator renormalization of physical OMEs up to four loops, 
hence we will not consider them further here.

\section{Feynman rules of alien operators}
\label{sec:feynman}
In this section we derive the Feynman rules of the alien operators. These were computed up to two loops in \cite{Hamberg:1991qt,Matiounine:1998ky,Blumlein:2022ndg}, and an extension to the three-loop level was recently presented in \cite{Gehrmann:2023ksf}. 
The Feynman rules for the gauge-invariant (physical) quark and gluon operators, up to the four-loop level, can be found e.g. in \cite{Falcioni:2022fdm,Gehrmann:2023ksf,Floratos:1977au,Floratos:1978ny,Mertig:1995ny,Kumano:1997qp,Hayashigaki:1997dn,Bierenbaum:2009mv,Klein:2009ig,Blumlein:2001ca,Velizhanin:2011es,Velizhanin:2014fua,Moch:2017uml,Moch:2021qrk,Falcioni:2023luc,Falcioni:2023vqq,Falcioni:2023tzp,Moch:2023tdj,Gehrmann:2023iah,Kniehl:2023bbk} and references therein. 
The generalization to arbitrary orders in perturbation theory is given in \cite{Somogyi:2024njx} \footnote{Note that  \cite{Somogyi:2024njx} also presents the corresponding rules for the operators with total derivatives, relevant for non-zero momentum flow through the operator vertex.}. 
We assume all momenta to be incoming and the total momentum flowing through the operator vertex to be zero, implying
\begin{equation}
    \sum_{i}p_i = 0.
\end{equation}

\subsection{Ghost operators}
The momenta of the ghost fields are taken to be $p_1$ and $p_2$, while $p_3,p_4,\dots$ denote the momenta of any additional gluons. 
As a check, we will compare our Feynman rules against the known ghost vertices with up to two additional gluons, which were computed in \cite{Gehrmann:2023ksf}. 
Because of different conventions for the operator definitions, the rules for the ghost vertices in the latter have to be divided by $i^N$. We can write the perturbative expansion of the ghost operator, cf.~Eq.~(\ref{eq:ghostFULL}), as
\begin{figure}[H]
\includegraphics[width=1\textwidth, trim = 72 705 72 72]{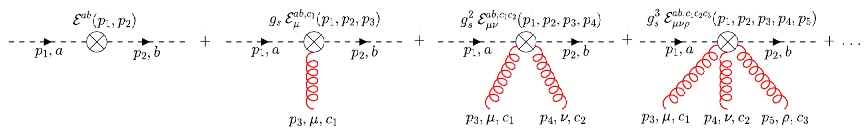}
\end{figure}
with
\begin{dmath}
\label{eq:ghost0}
    \varepsilon^{ab} = \frac{1+(-1)^{N}}{2}i^{N}\eta(N)\delta^{ab}(\Delta\cdot p_1)^{N},
\end{dmath}
\begin{dmath}
\label{eq:ghost1}
    \varepsilon^{ab,c_1}_{\mu} = \frac{1+(-1)^{N}}{2}i^{N-1}\Delta_{\mu}\,f^{a c_1 b}\sum_{\substack{i+j\\=N-3}}\eta_{ij}(\Delta\cdot p_1)(\Delta\cdot p_3)^{i}(\Delta\cdot p_2)^{j+1},
\end{dmath}
\begin{dmath}
\label{eq:ghost2}
    \varepsilon^{ab,c_1 c_2}_{\mu\nu}(p_1,p_2,p_3,p_4) = \frac{1+(-1)^{N}}{2}i^N\Delta_{\mu}\Delta_{\nu}\Bigg\{(f\:f)^{a c_1 c_2 b}\sum_{\substack{i+j+k\\=N-4}}\eta_{ijk}^{(1)}(\Delta\cdot p_{1})(\Delta\cdot p_{3})^{i}(\Delta\cdot p_{4})^{j}(\Delta\cdot p_{2})^{k+1}  +d_{4}^{a c_1 c_2 b}\sum_{\substack{i+j+k\\=N-4}}\eta_{ijk}^{(2)}(\Delta\cdot p_{1})(\Delta\cdot p_{3})^{i}(\Delta\cdot p_{4})^{j}(\Delta\cdot p_{2})^{k+1} +d_{\widehat{4ff}}^{a c_1 c_2 b}\sum_{\substack{i+j+k\\=N-4}}\eta_{ijk}^{(3)}(\Delta\cdot p_{1})(\Delta\cdot p_{3})^{i}(\Delta\cdot p_{4})^{j}(\Delta\cdot p_{2})^{k+1}\Bigg\} + [(p_{3},\mu,c_1)\text{$\leftrightarrow$} (p_{4},\nu,c_2)],
\end{dmath}
\begin{dmath}
\label{eq:ghost3}
    \varepsilon^{ab,c_1 c_2 c_3}_{\mu\nu\rho}(p_1,p_2,p_3,p_4,p_5) = -\frac{1+(-1)^{N}}{2}i^{N-1}\Delta_{\mu}\Delta_{\nu}\Delta_{\rho}\Bigg\{(f\:f\:f)^{a c_1 c_2 c_3 b}\sum_{\substack{i+j+k+l\\=N-5}}\eta_{ijkl}^{(1)}(\Delta\cdot p_{1})(\Delta\cdot p_{3})^{i}(\Delta\cdot p_{4})^{j}\times(\Delta\cdot p_{5})^{k}(\Delta\cdot p_{2})^{l+1}+d_{4f}^{a c_1 c_2 c_3 b}\sum_{\substack{i+j+k+l\\=N-5}}\eta_{ijkl}^{(2a)}(\Delta\cdot p_{1})(\Delta\cdot p_{3})^{i}(\Delta\cdot p_{4})^{j}(\Delta\cdot p_{5})^{k}(\Delta\cdot p_{2})^{l+1}+d_{4f}^{a b c_1 c_2 c_3}\sum_{\substack{i+j+k+l\\=N-5}}\eta_{ijkl}^{(2b)}(\Delta\cdot p_{1})(\Delta\cdot p_{3})^{i}(\Delta\cdot p_{4})^{j}(\Delta\cdot p_{5})^{k}(\Delta\cdot p_{2})^{l+1}\Bigg\} + \text{\: permutations}
\end{dmath}
where the `+ permutations' in the $O(g_s^3)$ rule in Eq.~(\ref{eq:ghost3}) denotes the fact that all permutations of the gluonic quantities (momenta, Lorentz and colour indices) have to be added. Note that $p_2$ in $\varepsilon^{ab}$ in Eq.~(\ref{eq:ghost0}) was eliminated using momentum conservation, $p_2=-p_1$. This then agrees with Eq.~(5.20) in \cite{Gehrmann:2023ksf} after dividing the latter by $i^N$, as discussed above. 
Similarly, after performing the summation, the $O(g_s)$ rule exactly matches Eq.~(5.21) in \cite{Gehrmann:2023ksf}. 
At $O(g_s^2)$, our expression for $\varepsilon^{ab,c_1 c_2}_{\mu\nu}(p_1,p_2,p_3,p_4)$ should be compared against Eq.~(5.22) in \cite{Gehrmann:2023ksf}. 
With $\eta_{ijk}^{(1)}$ given by Eq.~(\ref{eq:eta1-final}) and $\eta_{ijk}^{(2)}$ by Eq.~(\ref{eq:eta2-fin}), we find exact agreement with that expression \footnote{The term in our expression proportional to $(f\:f)^{a c_1 c_2 b}$ should be compared to the $f^{a_1 a_3 a}f^{a_2 a_4 a}$ part of Eq.~(5.22)  in \cite{Gehrmann:2023ksf} while our $d_{4}^{a c_1 c_2 b}$ rule should be compared to the one proportional to $d_{4}^{a_1 a_2 a_3 a_4}/C_A$.}. 
Finally the Feynman rule for the $d_{\widehat{4ff}}$ part of the ghost operator is computed using
\begin{dmath}
    \eta_{ijk}^{(3)} = 2\Bigg\{(c_{1}-c_{2})(-1)^{i+k}\Bigg[\binom{i+k+1}{i}-\binom{i+k+1}{k}\Bigg]+(-1)^{j+k+1}\Bigg[(c_{1}+2c_{2})\binom{j+k+1}{j}+(2c_{1}+c_{2})\binom{j+k+1}{k}\Bigg]+\binom{i+j+1}{i}\Bigg[(2c_{1}+c_{2})(-1)^{i+j}-c_{1}\binom{N-2}{k+1}\Bigg]+\binom{i+j+1}{j}\Bigg[c_{1}(-1)^{i+j}+2c_{2}(-1)^{i+j}-2c_{1}\binom{N-2}{k}-c_{1}\binom{N-2}{k+1}-c_{2}\binom{N-1}{k+1}\Bigg]+\binom{i+k+1}{i}\Bigg[2c_{1}\binom{N-2}{j}+c_{2}\binom{N-1}{j+1}\Bigg]-\binom{i+k+1}{k}\Bigg[2c_{1}\binom{N-2}{j}+c_{2}\binom{N-1}{j+1}\Bigg]+\binom{j+k+1}{j}\Bigg[2c_{1}\binom{N-2}{i}+c_{1}\binom{N-2}{i+1}+c_{2}\binom{N-1}{i+1}\Bigg]+c_{1}\binom{N-2}{i+1}\binom{j+k+1}{k}\Bigg\}
\end{dmath}
which follows from Eqs.~(\ref{eq:kappa3-gbrst}) and (\ref{eq:kappa3-fin}). 
As discussed in Sec.~\ref{sec:kappa3-ijk}, the free parameters $c_1,c_2$ can be determined by a computation of fixed-$N$ OMEs.

\subsection{Alien gluon operators}
Next we derive the Feynman rules for the gluonic EOM operator, whose perturbative expansion can be written as
\begin{figure}[H]
\includegraphics[width=1\textwidth, trim = 72 705 72 72]{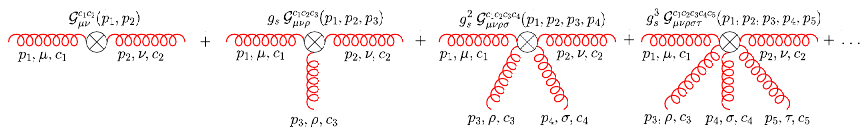}
\end{figure}
with
\begin{dmath}
\label{eq:gluon2}
  \mathcal{G}_{\mu\nu}^{c_1 c_2}(p_1,p_2) = \frac{1+(-1)^{N}}{2}i^N\eta(N)\delta^{c_1 c_2}(\Delta\cdot p_1)^{N-2}[2p_1^2\Delta_{\mu}\Delta_{\nu}-(\Delta\cdot p_1)(\Delta_{\mu}p_{1\nu}+\Delta_{\nu}p_{1\mu})],
\end{dmath}
\begin{dmath}
\label{eq:gluon3}
   \mathcal{G}_{\mu\nu\rho}^{c_1 c_2 c_3}(p_1,p_2,p_3) = -\frac{1+(-1)^{N}}{2}i^{N-1}f^{c_1 c_2 c_3}\Bigg\{\eta(N)(\Delta\cdot p_1)^{N-2}\Delta_{\mu}[p_{3\nu}\Delta_{\rho}-g_{\nu\rho}(\Delta\cdot p_3)+\Delta_{\rho}(p_2+p_3)_{\nu}]+\Delta_{\nu}\Delta_{\rho}[p_1^2\Delta_{\mu}-p_{1\mu}(\Delta\cdot p_1)]\sum_{\substack{i+j\\=N-3}}\kappa_{ij}(\Delta\cdot p_2)^i (\Delta\cdot p_3)^j \Bigg\}+ \text{permutations},
\end{dmath}
\begin{dmath}
\label{eq:gluon4}
  \mathcal{G}_{\mu\nu\rho\sigma}^{c_1 c_2 c_3 c_4}(p_1,p_2,p_3,p_4) = \frac{1+(-1)^{N}}{2}i^{N-2}f^{c_1 c_2 x}f^{x c_3 c_4}\Bigg\{\big[\Delta_{\nu}\Delta_{\rho}\Delta_{\sigma}(p_1+2p_2)_{\mu}-g_{\mu\nu}\Delta_{\rho}\Delta_{\sigma}(\Delta\cdot p_2)\big]\sum_{\substack{i+j\\=N-3}}\kappa_{ij}(\Delta\cdot p_3)^{i}(\Delta\cdot p_4)^{j}-g_{\nu\rho}\Delta_{\mu}\Delta_{\sigma}(\Delta\cdot p_1)^{N-2}+\big[p_1^2\Delta_{\mu}-p_{1\mu}(\Delta\cdot p_1)\big]\Delta_{\nu}\Delta_{\rho}\Delta_{\sigma}\sum_{\substack{i+j+k\\=N-4}}\kappa^{(1)}_{ijk}(\Delta\cdot p_2)^{i}(\Delta\cdot p_3)^{j}(\Delta\cdot p_4)^{j}\Bigg\}+\frac{1+(-1)^{N}}{2}\big[p_1^2\Delta_{\mu}-p_{1\mu}(\Delta\cdot p_1)\big]\Delta_{\nu}\Delta_{\rho}\Delta_{\sigma}\Bigg\{d_4^{c_1 c_2 c_3 c_4}\sum_{\substack{i+j+k\\=N-4}}\kappa^{(2)}_{ijk}(\Delta\cdot p_2)^{i}(\Delta\cdot p_3)^{j}(\Delta\cdot p_4)^{j}+d_{\widehat{4ff}}^{c_1 c_2 c_3 c_4}\sum_{\substack{i+j+k\\=N-4}}\kappa^{(3)}_{ijk}(\Delta\cdot p_2)^{i}(\Delta\cdot p_3)^{j}(\Delta\cdot p_4)^{j}\Bigg\}+ \text{permutations},
\end{dmath}
\begin{dmath}
\label{eq:gluon5}
  \mathcal{G}_{\mu\nu\rho\sigma\tau}^{c_1 c_2 c_3 c_4 c_5}(p_1,p_2,p_3,p_4,p_5) = \frac{1+(-1)^{N}}{2}i^{N-1}f^{c_1 c_2 x}f^{x c_3 y}f^{y c_4 c_5}\Bigg\{-g_{\mu\rho}\Delta_{\nu}\Delta_{\sigma}\Delta_{\tau}\sum_{\substack{i+j\\=N-3}}\kappa_{ij}(\Delta\cdot p_4)^{i}(\Delta\cdot p_5)^{j}+\Delta_{\rho}\Delta_{\sigma}\Delta_{\tau}\big[(p_1+2p_2)_{\mu}\Delta_{\nu}-(\Delta\cdot p_2)g_{\mu\nu}\big]\sum_{\substack{i+j+k\\=N-4}}\kappa_{ijk}^{(1)}(\Delta\cdot p_3)^{i}(\Delta\cdot p_4)^{j}(\Delta\cdot p_5)^{k}+\big[p_1^2\Delta_{\mu}-p_{1\mu}(\Delta\cdot p_1)\big]\Delta_{\nu}\Delta_{\rho}\Delta_{\sigma}\Delta_{\tau}\sum_{\substack{i+j+k+l\\=N-5}}\kappa_{ijkl}^{(1)}(\Delta\cdot p_2)^{i}(\Delta\cdot p_3)^{j}(\Delta\cdot p_4)^{k}(\Delta\cdot p_5)^{l}\Bigg\}+\frac{1+(-1)^{N}}{2}i^{N-1}d_{4f}^{c_1 c_2 c_3 c_4 c_5}\Bigg\{\Delta_{\mu}\Delta_{\nu}\Delta_{\rho}\big[(p_4+2p_5)_{\sigma}\Delta_{\tau}-(\Delta\cdot p_5)g_{\sigma\tau}\big]\sum_{\substack{i+j+k\\=N-4}}\kappa_{ijk}^{(2)}(\Delta\cdot p_1)^{i}(\Delta\cdot p_2)^{j}(\Delta\cdot p_3)^{k}+\big[p_1^2\Delta_{\mu}-p_{1\mu}(\Delta\cdot p_1)\big]\Delta_{\nu}\Delta_{\rho}\Delta_{\sigma}\Delta_{\tau}\sum_{\substack{i+j+k+l\\=N-5}}\kappa_{ijkl}^{(2)}(\Delta\cdot p_2)^{i}(\Delta\cdot p_3)^{j}(\Delta\cdot p_4)^{k}(\Delta\cdot p_5)^{l}\Bigg\}+ \text{permutations},
\end{dmath}
where again all permutations of gluon momenta, Lorentz and colour indices have to be added, if indicated by `+ permutations'. Note that $p_2$ in $\mathcal{G}_{\mu\nu}^{c_1 c_2}(p_1,p_2)$ in Eq.~(\ref{eq:gluon2}) was again eliminated using momentum conservation. This then agrees with Eq.~(5.23) in \cite{Gehrmann:2023ksf} and Eq.~(243) in \cite{Blumlein:2022ndg} after dividing the latter rules by $i^N$ to match to our conventions. 

For the $O(g_s)$ EOM vertex three contributions need to be taken into account,
\begin{itemize}
    \item the non-Abelian part of the field strength in the class I operator with $D\rightarrow\partial$,
    \item the $O(g_s)$ part of the covariant derivative acting on the Abelian part of the field strength in the class I operator and
    \item the class II operator, cf.~Eq.~(\ref{eq:eomII}), with $D\rightarrow\partial$ and keeping only the Abelian part of the field strength.
\end{itemize}
Our result matches the corresponding rules in the literature, cf. Eq.~(5.24) in \cite{Gehrmann:2023ksf} and Eq.~(244) in \cite{Blumlein:2022ndg} respectively (again after dividing by the overall $i^{N}$). 

Next the four-gluon vertex gets four contributions,
\begin{itemize}
    \item the $O(g_s)$ part of the covariant derivative acting on the non-Abelian part of the field strength in the class I operator,
    \item the non-Abelian part of the field strength in the class II operator with $D\rightarrow\partial$,
    \item the $O(g_s)$ part of the covariant derivative acting on the Abelian part of the field strength in the class II operator and
    \item the class III operator, cf.~Eq.~(\ref{eq:eomIII}), with $D\rightarrow\partial$ and keeping only the Abelian part of the field strength.
\end{itemize}
The second and third contributions depend on the lower-order coupling $\kappa_{ij}$, while the fourth one is written in terms of the couplings $\kappa_{ijk}^{(1)}$, $\kappa_{ijk}^{(2)}$ and $\kappa_{ijk}^{(3)}$ given by Eqs.~(\ref{eq:kappa1-final}), (\ref{eq:kappa2-fin}) and (\ref{eq:kappa3-fin}) respectively. The $(f\:f)$ and $d_4$ parts of our rule agree with Eq.~(5.25) in \cite{Gehrmann:2023ksf} \footnote{Note however that our result proportional to $d_4$ needs to be multiplied by a symmetry factor of $1/4!$ to match Eq.~(5.25) in \cite{Gehrmann:2023ksf}, which is again a consequence of differing conventions.}, while the $d_{\widehat{4ff}}$ part is new. 

Finally, as a new result \footnote{The corresponding result within the framework of ref.~\cite{Gehrmann:2023ksf} was recently announced in a conference talk \cite{Gehrmann:2024ggw}.}, we consider the five-gluon vertex 
  $\mathcal{G}_{\mu\nu\rho\sigma\tau}^{c_1 c_2 c_3 c_4 c_5}(p_1,p_2,p_3,p_4,p_5)$ 
in Eq.~(\ref{eq:gluon5}). 
Again we need to take into account higher-order contributions of the lower-point vertices. In particular, the $(f\:f\:f)$ part of the five-gluon rule gets four contributions,
\begin{itemize}
    \item the $O(g_s)$ part of the covariant derivative acting on the non-Abelian part of the field strength in the class II operator,
    \item the non-Abelian part of the field strength in the class III operator with $D\rightarrow\partial$,
    \item the $O(g_s)$ part of the covariant derivative acting on the Abelian part of the field strength in the class III operator,
    \item the class IV operator, cf.~Eq.~(\ref{eq:eomIV}), with $D\rightarrow\partial$ and keeping only the Abelian part of the field strength.
\end{itemize}
On the other hand the $d_{4f}$ part only gets three contributions,
\begin{itemize}
    \item the non-Abelian part of the field strength in the class III operator with $D\rightarrow\partial$,
    \item the $O(g_s)$ part of the covariant derivative acting on the Abelian part of the field strength in the class III operator and
    \item the class IV operator, cf.~Eq.~(\ref{eq:eomIV}), with $D\rightarrow\partial$ and keeping only the Abelian part of the field strength.
\end{itemize}

\subsection{Alien quark operators}
Finally in this section we provide the Feynman rules for the alien quark operators presented in Eqs.~(\ref{eq:eomI})-(\ref{eq:eomIII}). As mentioned above, these operators are written in terms of the same couplings as those in the gluon EOM operators. Assuming the momenta of the external quark fields to be $p_1$ and $p_2$ we have the following perturbative expansion
\begin{figure}[H]
\includegraphics[width=1\textwidth, trim = 72 705 72 72]{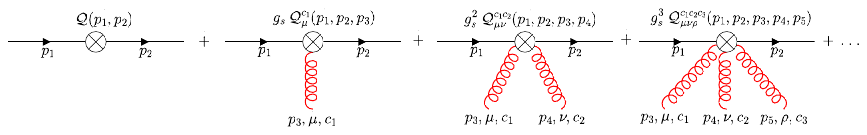}
\end{figure}
with
\begin{dmath}
    \mathcal{Q}(p_1,p_2) = 0,
\end{dmath}
\begin{dmath}
    \mathcal{Q}_{\:\:\mu}^{c_1}(p_1,p_2,p_3) = -\frac{1+(-1)^{N}}{2}i^{N}\eta(N)T^{c_1}\Delta_{\mu}\slashed\Delta(\Delta\cdot p_3)^{N-2},
\end{dmath}
\begin{dmath}
\label{eq:Qc1c2}
    \mathcal{Q}_{\:\:\mu\nu}^{c_1 c_2}(p_1,p_2,p_3,p_4) = [1+(-1)^N]i^{N-1}T^{a}f^{a c_1 c_2}\Delta_{\mu}\Delta_{\nu}\slashed\Delta\sum_{\substack{i+j\\=N-3}}\kappa_{ij}^{(1)}(\Delta\cdot p_3)^{i}(\Delta\cdot p_4)^{j},
\end{dmath}
\begin{dmath}
\label{eq:Qc1c2c3}
    \mathcal{Q}_{\:\:\mu\nu\rho}^{c_1 c_2 c_3}(p_1,p_2,p_3,p_4,p_5) =  [1+(-1)^N]i^{N}T^{a}\Delta_{\mu}\Delta_{\nu}\Delta_{\rho}\slashed\Delta\Bigg\{f^{a c_1 x}f^{c_2 c_3 x}\sum_{\substack{i+j+k\\=N-4}}\kappa_{ijk}^{(1)}(\Delta\cdot p_3)^{i}(\Delta\cdot p_4)^{j}(\Delta\cdot p_5)^{k}
    + d_{4}^{a c_1 c_2 c_3}\sum_{\substack{i+j+k\\=N-4}}\kappa_{ijk}^{(2)}(\Delta\cdot p_3)^{i}(\Delta\cdot p_4)^{j}(\Delta\cdot p_5)^{k} + d_{\widehat{4ff}}^{a c_1 c_2 c_3}\sum_{\substack{i+j+k\\=N-4}}\kappa_{ijk}^{(3)}(\Delta\cdot p_3)^{i}(\Delta\cdot p_4)^{j}(\Delta\cdot p_5)^{k}\Bigg\}+ [(p_3,\mu,c_1)\text{$\leftrightarrow$}(p_4,\nu,c_2)] + [(p_3,\mu,c_1)\text{$\rightarrow$}(p_5,\rho,c_3)\text{$\rightarrow$}(p_4,\nu,c_2)\text{$\rightarrow$}(p_3,\mu,c_1)]
\end{dmath}
The vertices with up to two additional gluons can be compared against the results presented in Eqs.~(5.17)-(5.19) of \cite{Gehrmann:2023ksf}. Dividing the latter by $i^N$ to match to our conventions, we find exact agreement. 
Note that Eqs.~(\ref{eq:Qc1c2})-(\ref{eq:Qc1c2c3}) contain an additional factor of two coming from the $[(p_4,\nu,c_2)\text{$\leftrightarrow$}(p_5,\rho,c_3)]$ permutation. This directly follows from the (anti-)symmetry properties of the $\kappa$-couplings, cf.~Eqs.~(\ref{eq:kappa1-antisym}), (\ref{eq:kappa2-sym}) and (\ref{eq:kappa3-sym}). 
Finally, because the $\kappa$-couplings enter the quark operator at one order in the strong coupling lower than in the gluon EOM one, we can push the perturbative expansion of the quark operator to one order higher. 
Consequently we also present the quark operator vertex at ${\cal O}(g_s^{4})$ with four additional gluons
\begin{figure}[H]
\centering
\includegraphics[width=0.3\textwidth, trim = 72 524 72 72]{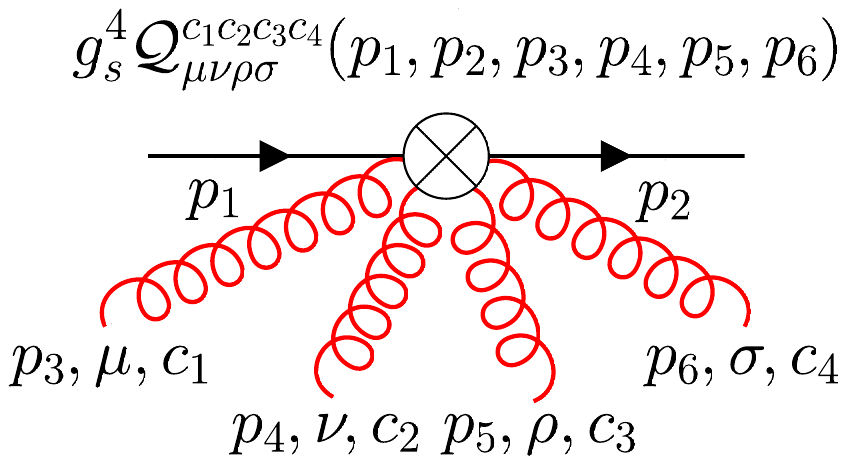}
\end{figure}
We find
\begin{dmath}
    \mathcal{Q}^{c_1 c_2 c_3 c_4}_{\mu\nu\rho\sigma}(p_1,p_2,p_3,p_4,p_5,p_6) = -\frac{1+(-1)^{N}}{2}i^{N-1}T^{a}\Delta_{\mu}\Delta_{\nu}\Delta_{\rho}\Delta_{\sigma}\slashed\Delta\Bigg\{(f\:f\:f)^{a c_1 c_2 c_3 c_4}\sum_{\substack{i+j+k+l\\=N-5}}\kappa_{ijkl}^{(1)}\times(\Delta\cdot p_{1})(\Delta\cdot p_{3})^{i}(\Delta\cdot p_{4})^{j}(\Delta\cdot p_{5})^{k}(\Delta\cdot p_{2})^{l+1}+d_{4f}^{a c_1 c_2 c_3 c_4}\sum_{\substack{i+j+k+l\\=N-5}}\kappa_{ijkl}^{(2)}(\Delta\cdot p_{1})(\Delta\cdot p_{3})^{i}(\Delta\cdot p_{4})^{j}(\Delta\cdot p_{5})^{k}(\Delta\cdot p_{2})^{l+1}\Bigg\} + \text{\: permutations}
\end{dmath}
with all permutations of the gluonic quantities (momenta, Lorentz and colour indices) to be added.

\section{Conclusions}
\label{sec:concl}

The kernels for parton evolution equations in QCD, i.e.\ splitting functions or the corresponding anomalous dimensions as their Mellin transforms, can be conveniently determined from the ultraviolet singularities of off-shell Green's functions with insertions of gauge-invariant twist-two spin-$N$ operators.
The renormalization of these OMEs, though, requires the computation of unphysical counterterms for the associated set of alien operators, which effectively describe vertices of two gluons, ghosts or quarks with any number $n\geq 0$ of additional gluons.
The couplings of these alien operators (EOM and ghost operators) are restricted by the fundamental symmetries, particularly the gBRST relations, which reflect the gauge theory characteristics of QCD.

The set of constraints for these couplings admits explicit solutions, valid for any spin $N$, which can be obtained using algorithms for symbolic summation to solve the recurrence relations.
A small number of boundary conditions in these solutions can be derived from the computation of the relevant OMEs at specific fixed values of $N$.
In addition, we have observed that the constraints contain a hierarchy, such that the function space of the couplings of alien operators with $n+1$ gluons can be derived from that of the $n$-gluon aliens. Thus, the basic ingredients in this bootstrap turn out to be the EOM and ghost operators with the smallest number of additional gluons at a given loop order.

We have provided results for all one-loop alien operator couplings needed in the renormalization of OMEs with physical (gauge-invariant) operators up to four loops, which represents the current frontier in splitting function computations.
This includes in particular the gluon EOM operator with five gluons attached, which is a new result. 
The all-$N$ solutions for the couplings that we have obtained can all be related to the fundamental one-loop counterterm $\eta(N)$ for the EOM and ghost operators of class I involving only two gluons or ghosts. 
We have also derived the corresponding Feynman rules and, whenever possible, compared them to those in the literature, finding full agreement.

A {\tt Mathematica} file with our results for the all-$N$ couplings necessary for the renormalization up to four loops is made available at the preprint server \url{https://arxiv.org}. We note that the expressions collected in this file have the fundamental one-loop counterterm $\eta(N)$ divided out.

The symmetries and the structure of the alien operators, that we have exploited in this study, are  independent of the order of perturbation theory.
Thus, we expect also analytic all-$N$ solutions beyond one loop for the couplings of the alien operators of class II and higher. 
We leave this task to future studies.

%
\subsection*{Acknowledgments}
\vspace*{-1mm}
The Feynman diagrams in this work are drawn using \textit{FeynGame}~\cite{Harlander:2020cyh,Harlander:2024qbn}. 

This work has been supported by the EU's Marie Sklodowska-Curie 
grant 101104792, {\it QCDchallenge};
the DFG through the Research Unit FOR 2926,
{\it Next Generation pQCD for Hadron Structure: Preparing for the EIC},
project number 40824754, DFG grant MO~1801/4-2,
the ERC Advanced Grant 101095857 {\it Conformal-EIC}; 
and by grant K143451 of the National Research, Development and Innovation Fund in Hungary.

{\footnotesize
\bibliographystyle{JHEP}
\bibliography{renormbib}
}

\end{document}